# Dual-support Smoothed Particle Hydrodynamics

Huilong Ren[a], Zili Dai[a], Xiaoying Zhuang[b,c,*], Timon Rabczuk[a,d,e,*]

[a]*Institute of Structural Mechanics, Bauhaus-Universitt Weimar, 99423 Weimar, Germany*
[b]*State Key Laboratory of Disaster Reduction in Civil Engineering, College of Civil Engineering,Tongji University, Shanghai 200092, China*
[c]*Institute of Conitnuum Mechanics, Leibniz University Hannover, Hannover, Germany*
[d]*Division of Computational Mechanics, Ton Duc Thang University, Ho Chi Minh City, Viet Nam*
[e]*Faculty of Civil Engineering, Ton Duc Thang University, Ho Chi Minh City, Viet Nam*

**Abstract**

In this paper we develop a dual-support smoothed particle hydrodynamics (DS-SPH) that naturally satisfies the conservation of momentum, angular momentum and energy when the varying smoothing length is utilized. The DS-SPH is based on the concept of dual-support, which is introduced to consider the unbalanced interactions between the particles with different smoothing lengths. Our DS-SPH formulation can be implemented in traditional SPH with little changes and improve the computational efficiency. Several numerical examples are presented to demonstrate the capability of the method.

## 1. Introduction

Smoothed particle hydrodynamics (SPH), which was invented in 1977 for modeling astrophysical problems [1, 2], has been one of the most popular mesh-free methods. The SPH method has attracted attentions of the researchers from a variety of fields since the beginning of the 1990s, and has been successfully applied to solve the problems, e.g. impact penetration in solids [3, 4, 5, 6], multiple-phase flows [7, 8, 9], free-surface flows in fluid dynamics [10, 11], Magnetohydrodynamics [12, 13, 14], solidification and phase-transition [15, 16], and so forth.

SPH is a Lagrangian method based on the kernel interpolation. In the SPH formulation, the partial differential equations are transformed into integral form with the kernel interpolation technique [17, 18]. The computational domain is represented by a set of particles which carry the physical properties (e.g. mass, density, velocity, position, pressure and internal energy). The particles move and properties change with time due to the interactions with neighboring particles.

One constraint in kernel interpolation is that the smoothing length is required to be constant for all particles. However, a computational efficient SPH implementation requires locally refined regions with areas of variable smoothing lengths - that is, with a roughly large $h$ to the sparse region and a relatively small $h$ to the clustered region. Such a strategy is expected to improve the accuracy of the solution at a relatively low computational cost. The similar situation could be found in FEM or FVM, where the dense nodes arrangement is used in domain of interest, while the coarse nodes arrangement in the other domain. In order to achieve the goal of the variable smoothing length, several methods have been proposed, e.g. averaged kernel method [19] and correction method with $\nabla h$ [20, 21, 22, 23]. The averaged kernel method uses the averaged smoothing lengths or averaged kernel functions, where the conservation of basic laws can not be guaranteed. For the other method, the gradient of smoothing length must be calculated to determine the optimal smoothing length, and a modified coefficient is introduced in all SPH formulations to preserve the conservations of basic laws (more details will be presented in §2). The optimal smoothing length is calculated iteratively, which makes the implementation of SPH more complicated. Therefore, it is desirable to find a new variable smoothing length method that is simple, efficient, and conserves all the basic laws, i.e. conservation of linear momentum, angular momentum and energy.

In this paper, we develop a new SPH formulation which naturally includes variable smoothing lengths. The new formulation named dual-support smoothed particle hydrodynamics (DS-SPH) is simple and intuitive, satisfies the conservation of momentum, angular momentum and energy when variable smoothing lengths are employed. The paper is outlined as follows. In §2, the theoretical background of SPH is reviewed. In §3, the concepts of support and dual-support are introduced, based on which we present a general method to reformulate the terms in classic SPH formulation. In §4, we convert the traditional SPH formulation into the dual-support SPH. In §5, the conservations of momentum, angular momentum and energy are proved. In §6, the issues related to the implementation of DS-SPH are discussed. In §7, four numerical examples are presented to verify the DS-SPH.

## 2. Theoretical background

*2.1. Kernel function*

Let $i$ denote the point in the global domain $\mathbf{\Omega}$; $\mathbf{r}_i$ is point $i$'s current coordinate. $H_i$ is the support for $i$ with a radius of $h_i$; the kernel function $W_i$ is defined:

$$W_{ij} := W_i(\mathbf{r}_{ij}, h_i) = w(\frac{r_{ij}}{h_i}) \tag{2.1}$$

$$\nabla_i W_{ij} := \frac{\partial W_{ij}}{\partial \mathbf{r}_i} = \frac{\partial W_{ij}}{\partial r_{ij}} \cdot \frac{\partial r_{ij}}{\partial \mathbf{r}_i} = \frac{\partial W_{ij}}{\partial r_{ij}} \cdot \frac{\mathbf{r}_{ij}}{r_{ij}} \tag{2.2}$$

$$\nabla_j W_{ij} := \frac{\partial W_{ij}}{\partial \mathbf{r}_j} = \frac{\partial W_{ij}}{\partial r_{ij}} \cdot \frac{\partial r_{ij}}{\partial \mathbf{r}_j} = \frac{\partial W_{ij}}{\partial r_{ij}} \cdot \frac{-\mathbf{r}_{ij}}{r_{ij}} \tag{2.3}$$

where $r_{ij} = \|\mathbf{r}_{ij}\|$, $\mathbf{r}_{ij} = \mathbf{r}_i - \mathbf{r}_j$, $w(\frac{r_{ij}}{h_i})$ is a scalar radial basis function.

When $h_i = h_j$, we have $W_{ij} = W_{ji}, \nabla_i W_{ij} = \nabla_i W_{ji}, \nabla_j W_{ij} = \nabla_j W_{ji}$. There are a wide variety of kernel functions, including the Gaussian kernel, cubic-spline kernel, quadratic kernel, quintic kernel, Wendland kernel [24]. More discussion on these kernel functions sees [18, 25].

*2.2. Kernel interpolation*

Kernel interpolation theory starts with the identity

$$f(\mathbf{r}_i) = \int f(\mathbf{r}_j) \delta(\mathbf{r}_i - \mathbf{r}_j) \mathrm{d}V_j \tag{2.4}$$

where $f$ is an arbitrary scalar variable and $\delta$ refers to the Dirac delta function. This integral is then approximated by replacing the delta function with a smoothing kernel $W$ with finite width $h_i$, e.g.

$$f(\mathbf{r}_i) = \int_{\mathbf{\Omega}} f(\mathbf{r}_j) W_{ij}(\mathbf{r}_i - \mathbf{r}_j, h_i) \mathrm{d}V_j = \int_{H_i} f(\mathbf{r}_j) W_{ij}(\mathbf{r}_i - \mathbf{r}_j, h_i) \mathrm{d}V_j \tag{2.5}$$

where $W$ has the property

$$\lim_{h_i \to 0} W(\mathbf{r}_i - \mathbf{r}_j, h_i) = \delta(\mathbf{r}_i - \mathbf{r}_j)$$

Using the integration by part and neglecting all surface terms, the derivative of $f(\mathbf{r}_i)$ is derived as

$$\nabla f(\mathbf{r}_i) = \int_{H_i} f(\mathbf{r}_j) \cdot \nabla_i W_{ij} \mathrm{d}V_j \tag{2.6}$$

Hence, the derivative of a function is transferred to the kernel function by kernel interpolation theory.

*2.3. Completeness of kernel function*

The completeness relates to the capability of kernel function to represent the rigid body modes and constant strain. According to [26], in the field of SPH, the completeness of the kernel function should satisfy

$$\sum_{j \in H_i} W_{ij} \mathbf{x}_j \Delta V_j = \mathbf{x}_i \tag{2.7}$$

The completeness of derivatives of kernel function should satisfy

$$\nabla \mathbf{u}_i = -\sum_{j \in H_i} \nabla_i W_{ij} \mathbf{u}_j \Delta V_j \tag{2.8}$$

The corrected derivatives method guarantees linear completeness of the derivative of a function [26]. According to reference [27], Eq.(2.8) is equivalent to

$$\sum_{j \in H_i} \mathbf{r}_{ij} \otimes \nabla_i W_{ij} \Delta V_j = \mathbf{I}, \text{ for } i = 1, ..., N. \tag{2.9}$$

$N$ is the number of particles. The kernel function in its continuous form fulfills zero- and first-order completeness and hence the 'standard' kernel function as given in Eq.(2.5) is sufficient. However, for discrete form, the completeness of kernel or kernel derivative is not guaranteed. There are many kernel correction methods, e.g. the symmetrization proposed by Monaghan [12], Randles & Libersky correction [4], Johnson & Beissel correction [28] and Krongauz-Belytschko Correction [29]. In the current paper, two simple correction methods are discussed.

The gradient correction of kernel function [27] in discrete form satisfying Eq.(2.9) is

$$\tilde{\nabla}_i W_{ij} = \mathbf{L}_i \nabla_i W_{ij} \qquad (2.10)$$
$$\mathbf{L}_i^{-1} = \sum_{j \in H_i} \Delta V_j \, \nabla_i W_{ij} \otimes (\mathbf{r}_j - \mathbf{r}_i),$$

where $\Delta V_j = m_j/\rho_j$. The gradient correction guarantees the conservation of angular momentum. Another simple correction method is mixed kernel and gradient correction, which fulfills zero- and first-order completeness. The zero-order (Shepard filter) is given by

$$\tilde{W}_{ij} = \frac{W_{ij}}{\sum_{H_i} W_{ij} \Delta V_j} \qquad (2.11)$$

Such correction is often adopted to reduce the pressure oscillations for particles near the boundary areas or close to free-surfaces when density summation is used instead of the continuity equation. Based on Eq.(2.11), the kernel gradient correction (or corrected derivatives method) is given by

$$\tilde{\nabla}_i \tilde{W}_{ij} = \mathbf{L}_i \nabla_i \tilde{W}_{ij} \qquad (2.12)$$
$$\mathbf{L}_i^{-1} = \sum_{j \in H_i} \Delta V_j \, \nabla_i \tilde{W}_{ij} \otimes (\mathbf{r}_j - \mathbf{r}_i)$$
$$\nabla_i \tilde{W}_{ij} = \frac{\nabla_i W_{ij} - \gamma(\mathbf{r}_i)}{\sum_{H_i} \Delta V_j W_{ij}}$$
$$\gamma(\mathbf{r}_i) = \frac{\sum_{H_i} \Delta V_j \nabla_i W_{ij}}{\sum_{H_i} \Delta V_j W_{ij}}$$

where $\Delta V_j = m_j/\rho_j$. It is not difficult to verify that both Eq.(2.10) and Eq.(2.12) satisfy Eq.(2.9).

The gradient correction Eq.(2.10) is much simpler than the so-called mixed correction Eq.(2.12). In the current paper, the gradient correction is preferred. The formulations with zero-ordercorrection or gradient kernel correction are straightforward in SPH. In other words, the replacement of $W_{ij} \rightarrow \tilde{W}_{ij}$ and $\nabla_i W_{ij} \rightarrow \tilde{\nabla}_i W_{ij}$ or $\nabla_i W_{ij} \rightarrow \tilde{\nabla}_i \tilde{W}_{ij}$ leads to the SPH formulation with corrected kernel function. Therefore, in this paper, we use the original form of kernel function to give the SPH formulation.

## 2.4. Variable smoothing length in SPH

There are several methods to deal with the variable smoothing length issue. The first method is implemented by averaging the kernel function [19] given by

$$\overline{W}_{ij} = \frac{1}{2}[W(|\mathbf{r}_{ij}|, h_i) + W(|\mathbf{r}_{ij}|, h_j)] \text{ or } \overline{W}_{ij} = W(|\mathbf{r}_{ij}|, [h_i + h_j]/2) \quad (2.13)$$

The averaged kernel between paired particles with different smoothing lengths, guarantees the anti-symmetrical pair-wised forces and thus preserves the symmetry of the particle interactions [18]. The second method is by considering the correction term $\nabla h$ [20, 21, 22, 23] in the equations of motion. The key idea is to relate the local number density of particles with the smoothing length and to keep the mass inside the smoothing sphere constant [22], i.e.

$$h(\mathbf{r}_i) \propto n(\mathbf{r}_i)^{-1/d}; \qquad n(\mathbf{r}_i) = \sum_{j \in H_i} W[(\mathbf{r}_i - \mathbf{r}_j), h(\mathbf{r}_i)] \quad (2.14)$$

The motion equation in [22] is

$$\frac{d\mathbf{v}_i}{dt} = -\sum_{j \in H_i} m_j \left[ f_i \frac{p_i}{\rho_i^2} \nabla_i W_{ij}(h_i) + f_j \frac{p_j}{\rho_j^2} \nabla_i W_{ij}(h_j) \right], \quad (2.15)$$

where $p_i$ is the pressure, the $f_i$ are defined by

$$f_i = \left[ 1 + \frac{h_i}{3\rho_i} \frac{\partial \rho_i}{\partial h_i} \right]^{-1}. \quad (2.16)$$

The quantities $\frac{\partial \rho_i}{\partial h_i}$ can be computed along with the densities themselves, more details please refer to [22].

Meanwhile, Monaghan [21] proposed a similar formulation to vary the smoothing length. In his work, the momentum equation is

$$\frac{d\mathbf{v}_i}{dt} = \sum_{j \in H_i} m_j \left[ \frac{p_i}{\Omega_i \rho_i^2} \nabla_i W_{ij}(h_i) + \frac{p_j}{\Omega_j \rho_j^2} \nabla_i W_{ji}(h_j) \right], \quad (2.17)$$

where $\Omega_i$ is given by

$$\Omega_i = 1 - \frac{\partial D_i}{\partial \rho_i} \sum_{j \in H_i} m_j \frac{\partial W_{ij}(h_i)}{\partial h_i}. \quad (2.18)$$

$D_i$ is a function to prevent arbitrarily large $h_i$ when $\rho_i$ becomes very small, which is given by

$$D_i = \frac{A}{1 + B\rho_i^{1/d}}, \quad (2.19)$$

where $A$ and $B$ are constants. More details please refer to [21]. The coefficient $\Omega_i$ also exists in continuity and energy equations.

Both the expressions in Eq.(2.16) and Eq.(2.18) are related to term $\nabla h_i$. The formulations in Eq.(2.15) and Eq.(2.17) are a little complicated due to the calculation of $\nabla h_i$. In fact, the basic idea behind is to find the proper value $h_i$ which satisfies Eq.(2.14). In order to find the desirable smoothing length, Eq.(2.14) is transformed into a set of two simultaneous equations which is computed at the location of particle $i$.

$$\rho(\mathbf{r}_i) = \sum_{j \in H_i} m_j W(\mathbf{r}_i - \mathbf{r}_j, h_i); \qquad h(\mathbf{r}_i) = \eta \left(\frac{m_i}{\rho_i}\right)^{1/d} \qquad (2.20)$$

where $\eta$ is a parameter specifying the smoothing length in units of the mean particle spacing $(m/\rho)^{1/d}$, $d$ is the number of dimensions. These two equations can be solved simultaneously using standard root-finding methods such as Newton-Raphson or Bisection [30]. However, the root-finding methods inevitably increase the computational cost and make the SPH implementation more complex.

## 3. Support domain and dual-support domain

In this section, the key concepts of support and dual-support are presented and the dual property of dual-support is proved. The new concept provides great flexibility to convert the traditional constant support SPH to dual-support SPH allowing for variable smoothing length for each particles.

### 3.1. support and dual-support

The conventional variable smoothing lengths SPH considers the unbalanced interaction with averaged kernel function, correction term $\nabla h$ or other methods. One basis of these methods is the single support domain. However, the single support domain cannot elegantly resolve the unbalanced interaction with different support radii. In the variable smoothing lengths SPH, one common situation as shown in Fig.1 is that $j \in H_i$, $i \notin H_j$, in other words, $j$ exerts force on $i$ while $i$ exerts no force on $j$. The unilateral force violates the Newton's third law. Therefore, a single variable support is not sufficient to define the interactions between particles and the new concept of support and dual-support is introduced subsequently. Note that the concept of dual-support is borrowed from the "dual-horizon" in peridynamics [31].

**Support**
The support $H_i$ for point $i$ is defined as a domain related to $i$. When the domain is centered at $i$ with a radius of $h_i$, the support $H_i$ can be given as

$$H_i = \{\mathbf{r}_j \mid \|\mathbf{r}_i - \mathbf{r}_j\| \leq h_i\} \qquad (3.1)$$

One example for variable support domain is shown in Fig. 2, where $\{\mathbf{r}_1, \mathbf{r}_2, \mathbf{r}_4, \mathbf{r}_6\} \in H_0$.
**Dual-support**

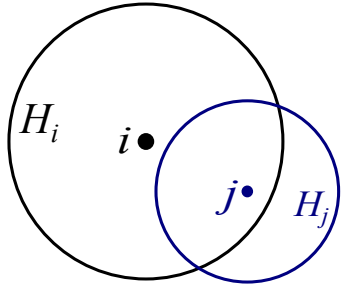

Figure 1: The location of two points with different support domains. $j \in H_i$, $i \notin H_j$

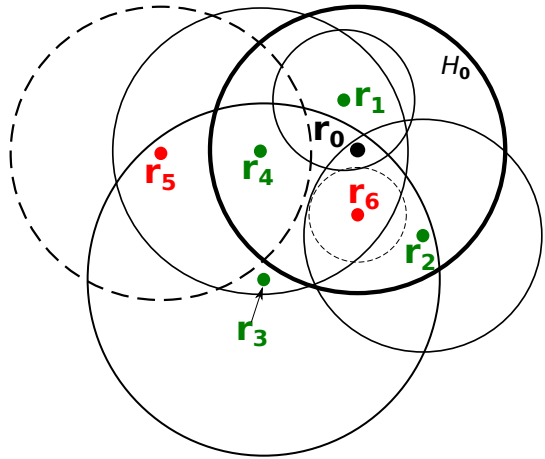

Figure 2: The schematic diagram for support and dual-support, all circles above are supports. The green points $\{\mathbf{r}_1, \mathbf{r}_2, \mathbf{r}_3, \mathbf{r}_4\} \in H'_0$, whose supports are denoted by thin solid line; the red points $\{\mathbf{r}_5, \mathbf{r}_6\} \notin H'_0$, whose supports are denoted by dashed line

Dual-support of $i$ is defined as a union of the points whose supports include $i$, denoted by

$$H'_i = \{\mathbf{r}_j \mid \mathbf{r}_i \in H_j\} \tag{3.2}$$

In the notation of dual-support $H'_i$, the superscript prime indicates dual, and the subscript $i$ denotes the object particle. When the support is defined as circle or sphere centered at that point, $i$'s dual-support can be expressed as

$$H'_i = \{\mathbf{r}_j \mid \|\mathbf{r}_i - \mathbf{r}_j\| \leq h_j\} \tag{3.3}$$

For any point $i$, the shape of $H'_i$ is arbitrary, and depends on the sizes and shapes of supports as well as the locations of the particles. For example, as shown in Fig. 2, the dual-support with respect to $\mathbf{r}_0$ contains particles $\mathbf{r}_1$, $\mathbf{r}_2$, $\mathbf{r}_3$ and $\mathbf{r}_4$, whose supports are denoted by thin solid circles. Particles $\mathbf{r}_5$ and $\mathbf{r}_6$ are not included in the dual-support of $\mathbf{r}_0$ since their supports do not include $\mathbf{r}_0$.

For models with constant supports, as $j \in H_i \Leftrightarrow i \in H_j$, $H'_i$ is equal to $H_i$ and therefore support and dual-support degenerate to the constant support in traditional SPH.

*3.2. The dual property of dual-support*

Let $\mathcal{F}(i,j)$ be any expression depend on two points $i, j$. The dual property of dual-support is that the double integral of the term $\mathcal{F}(i,j)$ in dual-support can be converted to the double integral of the term $\mathcal{F}(j,i)$ in support, as shown in Eq. (3.4) and Eq. (3.5). The key idea lies in that the term $\mathcal{F}(i,j)$ can be both interpreted in $H_i$ and $H'_j$. For the sake of simplicity, the proof of dual property of dual-support is given in Appendix A.

$$\sum_{i \in \Omega} \left( \sum_{j \in H'_i} \mathcal{F}(i,j) \Delta V_j \right) \Delta V_i = \sum_{i \in \Omega} \left( \sum_{j \in H_i} \mathcal{F}(j,i) \Delta V_j \right) \Delta V_i \qquad \text{discrete form} \tag{3.4}$$

$$\int_{i \in \Omega} \int_{j \in H'_i} \mathcal{F}(i,j) \,\mathrm{d}V_j \mathrm{d}V_i = \int_{i \in \Omega} \int_{j \in H_i} \mathcal{F}(j,i) \,\mathrm{d}V_j \mathrm{d}V_i \qquad \text{continuous form} \tag{3.5}$$

*3.3. The dual formulation based on support and dual-support*

Let $\rho$ be any scalar, $f$ be any scalar field, the SPH formulation of $\frac{\nabla f_i}{\rho_i}$ can be conventionally obtained as

$$\begin{aligned}\frac{\nabla f_i}{\rho_i} &= \nabla_i(\frac{f}{\rho}) + \frac{f_i}{\rho_i^2} \nabla_i \rho \\ &= \int_{H_i} \frac{f_j}{\rho_j^2} \nabla_i W_{ij} \rho_j \mathrm{d}V_j + \int_{H_i} \frac{f_i}{\rho_i^2} \nabla_i W_{ij} \rho_j \mathrm{d}V_j \end{aligned} \tag{3.6}$$

In order to get the dual formulation, we perform Eq. (3.7) in the term with $f_j$ of Eq. (3.6)

$$H_i \to H'_i, \qquad \nabla_i W_{ij} \to -\nabla_j W_{ji} \tag{3.7}$$

Then, we get the dual-support formulation

$$\frac{\nabla f_i}{\rho_i} \to -\int_{H'_i} \frac{f_j}{\rho_j^2} \nabla_j W_{ji} \rho_j \mathrm{d}V_j + \int_{H_i} \frac{f_i}{\rho_i^2} \nabla_i W_{ij} \rho_j \mathrm{d}V_j \tag{3.8}$$

Note that for points with identical smoothing length, $h_i = h_j$ leads to $H_i = H'_i$ and $\nabla_i W_{ij} = -\nabla_j W_{ji}$ in Eq. (3.7). Eq. (3.7) forms the key step to convert traditional constant support SPH to dual-support SPH. With the operation given by Eq. (3.7), let $\phi$ be any scalar, $\mathbf{A}$ be any vector, some dual-support formulations include

$$\phi_i \nabla_i \times \mathbf{A} = \nabla_i \times (\phi \mathbf{A}) + \mathbf{A} \times \nabla_i \phi$$
$$\to \int_{H'_i} \phi_j \mathbf{A}_j \times \nabla_j W_{ji} \mathrm{d}V_j + \int_{H_i} \phi_j \mathbf{A}_i \times \nabla_i W_{ij} \mathrm{d}V_j \tag{3.9}$$

$$\frac{1}{\phi_i} \nabla_i \times \mathbf{A} = \nabla_i \times (\frac{\mathbf{A}}{\phi}) - \frac{\mathbf{A} \times \nabla_i \phi}{\phi^2}$$
$$\to \int_{H'_i} \frac{\mathbf{A}_j}{\phi_j^2} \times \nabla_j W_{ji} \phi_j \mathrm{d}V_j - \int_{H_i} \frac{\mathbf{A}_i}{\phi_i^2} \times \nabla_i W_{ij} \phi_j \mathrm{d}V_j \tag{3.10}$$

The second order derivative can be obtained in a similar way. One expression recommended by [32, 33] for second derivative is

$$(\frac{1}{\rho} \nabla \cdot (\mu \nabla \mathbf{v}))_i = \sum_{j \in H_i} m_j \frac{\mu_j + \mu_i}{\rho_i \rho_j} \frac{\mathbf{r}_{ij} \cdot \nabla_i W_{ij}}{r_{ij}^2 + \eta^2} \mathbf{v}_{ij} \tag{3.11}$$

$$\to \sum_{j \in H_i} m_j \frac{\mu_i}{\rho_i \rho_j} \frac{\mathbf{r}_{ij} \cdot \nabla_i W_{ij}}{r_{ij}^2 + \eta^2} \mathbf{v}_{ij} - \sum_{j \in H'_i} m_j \frac{\mu_j}{\rho_i \rho_j} \frac{\mathbf{r}_{ji} \cdot \nabla_j W_{ji}}{r_{ji}^2 + \eta^2} \mathbf{v}_{ji} \tag{3.12}$$

which requires only first spatial derivatives, where $\mathbf{v}_{ij} = \mathbf{v}_i - \mathbf{v}_j$, $\eta$ is a small number introduced to avoid a zero denominator during computations and is set to $0.1h$. Eq.(3.11) is the combination of SPH formulation of first derivative and first order finite difference.

And for temperature:

$$(\frac{1}{\rho} \nabla \cdot (\kappa \nabla T))_i = \sum_{j \in H_i} m_j \frac{\kappa_i + \kappa_j}{\rho_i^2} \frac{\mathbf{r}_{ij} \cdot \nabla_i W_{ij}}{r_{ij}^2 + \eta^2} T_{ij}$$

$$\to -\sum_{j \in H'_i} m_j \frac{\kappa_j}{\rho_j^2} \frac{\mathbf{r}_{ji} \cdot \nabla_j W_{ji}}{r_{ji}^2 + \eta^2} T_{ji} + \sum_{j \in H_i} m_j \frac{\kappa_i}{\rho_i^2} \frac{\mathbf{r}_{ij} \cdot \nabla_i W_{ij}}{r_{ij}^2 + \eta^2} T_{ij} \tag{3.13}$$

where $\kappa_i$ is thermal conductivity associated to particle $i$, $T$ denotes the temperature, and $T_{ij} = T_i - T_j$.

## 4. Dual-support smoothed particle hydrodynamics

In this paper, we apply the dual-support formulation to fluids. The Lagrangian form of the fluid dynamics equations is given as

$$\frac{d\rho}{dt} = -\rho \nabla \cdot \mathbf{v} \qquad \text{continuity equation} \qquad (4.1)$$

$$\frac{d\mathbf{v}}{dt} = \frac{1}{\rho} \nabla \cdot \boldsymbol{\sigma} + \mathbf{b} \qquad \text{linear momentum} \qquad (4.2)$$

$$\frac{de}{dt} = \frac{1}{\rho} \nabla \cdot (\boldsymbol{\sigma} \cdot \mathbf{v}) - \frac{1}{\rho} \nabla \cdot \mathbf{q} + \mathbf{b} \cdot \mathbf{v} \qquad \text{specific energy} \qquad (4.3)$$

$$\frac{d\epsilon}{dt} = \frac{1}{\rho} \boldsymbol{\sigma} : \nabla \mathbf{v} - \frac{1}{\rho} \nabla \cdot \mathbf{q} \qquad \text{internal energy} \qquad (4.4)$$

where $\frac{d}{dt}$ is the material time derivative, $\boldsymbol{\sigma}$ is the Cauchy stress; $\mathbf{q} = -\kappa \nabla T$ is the heat flux density vector, $\kappa$ is the thermal conductivity and $T$ denotes the temperature; $\mathbf{b}$ is the body force density, $\epsilon$ is the internal energy density, $e = \frac{v^2}{2} + \epsilon$ is the specific energy per unit mass. For magnetic fluid, $e = \frac{1}{2}v^2 + \epsilon + \frac{1}{2\mu_0} B^2/\rho$, where $B = |\mathbf{B}|$ and $\mathbf{B}$ denotes the magnetic field. Some constitutions for fluid include, for examples,

$$\boldsymbol{\sigma} = -p\mathbf{I} \qquad \text{inviscid fluid} \qquad (4.5)$$

$$\boldsymbol{\sigma} = (-p + \lambda \nabla \cdot \mathbf{v})\mathbf{I} + \mu(\nabla \otimes \mathbf{v} + \mathbf{v} \otimes \nabla) \qquad \text{newtonian fluid} \qquad (4.6)$$

$$\boldsymbol{\sigma} = -(p + \frac{1}{2\mu_0} B^2)\mathbf{I} + \frac{1}{\mu_0} \mathbf{B} \otimes \mathbf{B} \qquad \text{magnetic inviscid fluid} \qquad (4.7)$$

where $p$ is the pressure; $\mu$ is the dynamic viscosity, and $\lambda$ is the second coefficient of viscosity [34], and $\boldsymbol{\sigma}$ is always symmetric.

In this paper, we omit the derivation for the governing equations and give the SPH formulations directly. The continuity equation is discretized into [18]

$$\rho_i = \sum_{j \in H_i} m_j \tilde{W}_{ij}, \qquad (4.8)$$

$$\frac{d\rho_i}{dt} = \sum_{j \in H_i} m_j \mathbf{v}_{ij} \nabla_i W_{ij}. \qquad (4.9)$$

In the Lagrangian formulation, the conservation of mass is naturally satisfied, therefore, we are not going to recast the continuity equation to dual-support formulation.

*4.1. General SPH formulation with dual-support*

Regardless of the material constitutions, the SPH formulation with constant smoothing length for equations of motion is

$$\frac{d\mathbf{v}_i}{dt} = \sum_{j \in H_i} m_j \left( \frac{\boldsymbol{\sigma}_i}{\rho_i^2} + \frac{\boldsymbol{\sigma}_j}{\rho_j^2} \right) \cdot \nabla_i W_{ij} + \mathbf{b}, \qquad (4.10)$$

and for energy equation is

$$\frac{\mathrm{d}e_i}{\mathrm{d}t} = \sum_{j \in H_i} m_j \left( \frac{\boldsymbol{\sigma}_i}{\rho_i^2} : (\mathbf{v}_j \otimes \nabla_i W_{ij}) + \frac{\boldsymbol{\sigma}_j}{\rho_j^2} : (\mathbf{v}_i \otimes \nabla_i W_{ij}) \right)$$
$$+ \sum_{j \in H_i} \frac{m_j}{\rho_i \rho_j} \frac{(\kappa_i + \kappa_j) T_{ij}}{r_{ij}^2 + \eta^2} \mathbf{r}_{ij} \cdot \nabla_i W_{ij} + \mathbf{b} \cdot \mathbf{v}_i. \tag{4.11}$$

(Eq.(4.11)-$\mathbf{v}_i \cdot$ Eq.(4.10)) leads to the energy equation based on internal energy

$$\frac{\mathrm{d}\epsilon_i}{\mathrm{d}t} = -\sum_{j \in H_i} m_j \frac{\boldsymbol{\sigma}_i}{\rho_i^2} : (\mathbf{v}_{ij} \otimes \nabla_i W_{ij}) + \sum_{j \in H_i} \frac{m_j}{\rho_i \rho_j} \frac{(\kappa_i + \kappa_j) T_{ij}}{r_{ij}^2 + \eta^2} \mathbf{r}_{ij} \cdot \nabla_i W_{ij}. \tag{4.12}$$

In order to allow variable smoothing lengths, performing Eq. (3.7) in Eq.(4.10), Eq.(4.11) and Eq.(4.12), the dual-support formulations for equations of motion and energy equations are

$$\frac{\mathrm{d}\mathbf{v}_i}{\mathrm{d}t} = \sum_{j \in H_i} m_j \frac{\boldsymbol{\sigma}_i}{\rho_i^2} \cdot \nabla_i W_{ij} - \sum_{j \in H_i'} m_j \frac{\boldsymbol{\sigma}_j}{\rho_j^2} \cdot \nabla_j W_{ji} + \mathbf{b} \tag{4.13}$$

$$\frac{\mathrm{d}e_i}{\mathrm{d}t} = \sum_{j \in H_i} m_j \frac{\boldsymbol{\sigma}_i}{\rho_i^2} : (\mathbf{v}_j \otimes \nabla_i W_{ij}) - \sum_{j \in H_i'} m_j \frac{\boldsymbol{\sigma}_j}{\rho_j^2} : (\mathbf{v}_i \otimes \nabla_j W_{ji}) +$$
$$\sum_{j \in H_i} \frac{m_j}{\rho_i \rho_j} \frac{\kappa_i T_{ij}}{r_{ij}^2 + \eta^2} \mathbf{r}_{ij} \cdot \nabla_i W_{ij} - \sum_{j \in H_i} \frac{m_j}{\rho_i \rho_j} \frac{\kappa_j T_{ji}}{r_{ji}^2 + \eta^2} \mathbf{r}_{ji} \cdot \nabla_j W_{ji} + \mathbf{b} \cdot \mathbf{v}_i \tag{4.14}$$

$$\frac{\mathrm{d}\epsilon_i}{\mathrm{d}t} = -\sum_{j \in H_i} m_j \frac{\boldsymbol{\sigma}_i}{\rho_i^2} : (\mathbf{v}_{ij} \otimes \nabla_i W_{ij}) +$$
$$\sum_{j \in H_i} \frac{m_j}{\rho_i \rho_j} \frac{\kappa_i T_{ij}}{r_{ij}^2 + \eta^2} \mathbf{r}_{ij} \cdot \nabla_i W_{ij} - \sum_{j \in H_i} \frac{m_j}{\rho_i \rho_j} \frac{\kappa_j T_{ji}}{r_{ji}^2 + \eta^2} \mathbf{r}_{ji} \cdot \nabla_j W_{ji}. \tag{4.15}$$

It should be mentioned that the constitutive relation is not specified. The application of dual-support SPH on solid mechanics is straightforward.

4.2. SPH formulations with artificial viscosity

When considering the pressure, artificial viscosity and physical viscosity in details, the SPH formulation for equations of motion [18, 35] is

$$\frac{\mathrm{d}\mathbf{v}_i}{\mathrm{d}t} = -\sum_{j \in H_i} m_j \left( \frac{p_i}{\rho_i^2} + \frac{p_j}{\rho_j^2} + \Pi_{ij} \right) \nabla_i W_{ij} + \mathbf{b}, \tag{4.16}$$

where $\Pi_{ij}$ is the artificial viscosity. The direction of artificial viscosity force is parallel with $\mathbf{r}_{ij}$. The artificial viscosity is often used in the momentum equation of SPH to model strong shocks and prevent particles from interpenetration [36, 17]. $\Pi_{ij}$ is given by

$$\Pi_{ij} = \begin{cases} \dfrac{-\alpha \bar{c}_{ij} \tilde{\mu}_{ij} + \beta \tilde{\mu}_{ij}^2}{\bar{\rho}_{ij}}, & \text{if } \mathbf{v}_{ij} \cdot \mathbf{r}_{ij} < 0; \\ 0, & \text{otherwise,} \end{cases} \quad (4.17)$$

where $\tilde{\mu}_{ij} = (\bar{h}_{ij} \mathbf{v}_{ij} \cdot \mathbf{r}_{ij})/(\mathbf{r}_{ij}^2 + \eta^2)$, $\bar{\rho}_{ij} = 0.5(\rho_i + \rho_j)$, $\bar{c}_{ij} = 0.5(c_i + c_j)$, $\bar{h}_{ij} = 0.5(h_i + h_j)$; $c_i$ is the sound speed associated to particle $i$, $\alpha$ and $\beta$ are constants that are all typically set around 1.0 [12].

The internal energy equation based on artificial viscosity is [37]

$$\frac{d\epsilon_i}{dt} = \frac{1}{2} \sum_{j \in H_i} m_j \left( \frac{p_i}{\rho_i^2} + \frac{p_j}{\rho_j^2} + \Pi_{ij} \right) \nabla_i W_{ij} \cdot \mathbf{v}_{ij} + \sum_{j \in H_i} \frac{m_j}{\rho_i \rho_j} \frac{(\kappa_i + \kappa_j) T_{ij}}{r_{ij}^2 + \eta^2} \mathbf{r}_{ij} \cdot \nabla_i W_{ij}. \quad (4.18)$$

Performing Eq. (3.7), the dual-support formulation of Eq.(4.16) is

$$\frac{d\mathbf{v}_i}{dt} = -\sum_{j \in H_i} m_j \left( \frac{p_i}{\rho_i^2} + \frac{\Pi_{ij}}{2} \right) \nabla_i W_{ij} + \sum_{j \in H_i'} m_j \left( \frac{p_j}{\rho_j^2} + \frac{\Pi_{ji}}{2} \right) \nabla_j W_{ji} + \mathbf{b}. \quad (4.19)$$

In Eq.(4.19), the direction of term $\nabla_j W_{ji}$ in $H_i'$ is $j \to i$, hence $\left( \frac{p_j}{\rho_j^2} + \frac{\Pi_{ji}}{2} \right) \nabla_j W_{ji}$ can be interpreted as the force density acted on $i$ due to $j$, where the unit of force density is force per volume squared. Similarly, the direction of term $\nabla_i W_{ij}$ in $H_i$ is $i \to j$, $\left( \frac{p_i}{\rho_i^2} + \frac{\Pi_{ij}}{2} \right) \nabla_i W_{ij}$ can be interpreted as the force density acted on $j$ due to $i$; as $i$ exerts a force on $j$, based on Newton's third law, $i$ will undertake a reaction force, so the negative symbol "$-$" in $H_i$ denotes the reaction force. Therefore, in the momentum equation, the expression in $H_i$ represents the reaction force density acted on $i$ as $i$ exerting a direct force on other particles, while the expression in $H_i'$ represents the direct force density acting on $i$ when another particle exerts force to $i$. Within the framework of dual-support SPH, any type of force can be divided into pairs, one is the direct force in dual-support domain, the other is the reaction force in support domain, which are always pair-wised with same magnitude but opposite direction.

The dual-support formulation for energy equation with artificial viscosity is

$$\begin{aligned}\frac{d\epsilon_i}{dt} =& \frac{1}{2} \sum_{j \in H_i} m_j \left( \frac{p_i}{\rho_i^2} + \frac{\Pi_{ij}}{2} \right) \nabla_i W_{ij} \cdot \mathbf{v}_{ij} + \frac{1}{2} \sum_{j \in H_i'} m_j \left( \frac{p_j}{\rho_j^2} + \frac{\Pi_{ji}}{2} \right) \nabla_j W_{ji} \cdot \mathbf{v}_{ji} \\ &+ \sum_{j \in H_i} \frac{m_j}{\rho_i^2} \frac{\kappa_i T_{ij}}{r_{ij}^2 + \eta^2} \mathbf{r}_{ij} \cdot \nabla_i W_{ij} - \sum_{j \in H_i'} \frac{m_j}{\rho_j^2} \frac{\kappa_j T_{ji}}{r_{ji}^2 + \eta^2} \mathbf{r}_{ji} \cdot \nabla_j W_{ji}.\end{aligned} \quad (4.20)$$

### 4.3. SPH formulation with physical viscosity

Another viscosity in fluid is the physical viscosity. The momentum equation with physical viscosity is

$$\frac{d\mathbf{v}_i}{dt} = -\sum_{j \in H_i} m_j \left(\frac{p_i}{\rho_i^2} + \frac{p_j}{\rho_j^2}\right) \nabla_i W_{ij} + \sum_{j \in H_i} m_j \frac{\mu_i + \mu_j}{\rho_i \rho_j} \frac{\mathbf{r}_{ij} \cdot \nabla_i W_{ij}}{r_{ij}^2 + \eta^2} \mathbf{v}_{ij} + \mathbf{b} \quad (4.21)$$

The internal energy equation based on physical viscosity is

$$\begin{aligned}\frac{d\epsilon_i}{dt} =& \frac{1}{2} \sum_{j \in H_i} m_j \left(\frac{p_i}{\rho_i^2} + \frac{p_j}{\rho_j^2}\right) \nabla_i W_{ij} \cdot \mathbf{v}_{ij} - \frac{1}{2} \sum_{j \in H_i} \frac{m_j(\mu_i + \mu_j)}{\rho_i \rho_j} \frac{\mathbf{r}_{ij} \cdot \nabla_i W_{ij}}{r_{ij}^2 + \eta^2} v_{ij}^2 \\ &+ \sum_{j \in H_i} \frac{m_j}{\rho_i \rho_j} \frac{(\kappa_i + \kappa_j) T_{ij}}{r_{ij}^2 + \eta^2} \mathbf{r}_{ij} \cdot \nabla_i W_{ij}\end{aligned} \quad (4.22)$$

So the dual-support formulation of Eq.(4.21) based on physical viscosity is

$$\begin{aligned}\frac{d\mathbf{v}_i}{dt} =& -\sum_{j \in H_i} m_j \frac{p_i}{\rho_i^2} \nabla_i W_{ij} + \sum_{j \in H_i'} m_j \frac{p_j}{\rho_j^2} \nabla_j W_{ji} \\ &- \sum_{j \in H_i'} m_j \frac{\mu_j}{\rho_j^2} \frac{\mathbf{r}_{ji} \cdot \nabla_j W_{ji}}{r_{ji}^2 + \eta^2} \mathbf{v}_{ji} + \sum_{j \in H_i} m_j \frac{\mu_i}{\rho_i^2} \frac{\mathbf{r}_{ij} \cdot \nabla_i W_{ij}}{r_{ij}^2 + \eta^2} \mathbf{v}_{ij} + \mathbf{b}\end{aligned} \quad (4.23)$$

The dual-support formulation of Eq.(4.22) is

$$\begin{aligned}\frac{d\epsilon_i}{dt} =& \frac{1}{2} \sum_{j \in H_i} m_j \frac{p_i}{\rho_i^2} \nabla_i W_{ij} \cdot \mathbf{v}_{ij} + \frac{1}{2} \sum_{j \in H_i'} m_j \frac{p_j}{\rho_j^2} \nabla_j W_{ji} \cdot \mathbf{v}_{ji} \\ &- \frac{1}{2} \sum_{j \in H_i} m_j \frac{\mu_i}{\rho_i^2} \frac{\mathbf{r}_{ij} \cdot \nabla_i W_{ij}}{r_{ij}^2 + \eta^2} v_{ij}^2 - \frac{1}{2} \sum_{j \in H_i'} m_j \frac{\mu_j}{\rho_j^2} \frac{\mathbf{r}_{ji} \cdot \nabla_j W_{ji}}{r_{ji}^2 + \eta^2} v_{ji}^2 \\ &+ \sum_{j \in H_i} m_j \frac{\kappa_i}{\rho_i^2} \frac{T_{ij}}{r_{ij}^2 + \eta^2} \mathbf{r}_{ij} \cdot \nabla_i W_{ij} - \sum_{j \in H_i'} m_j \frac{\kappa_j}{\rho_j^2} \frac{T_{ji}}{r_{ji}^2 + \eta^2} \mathbf{r}_{ji} \cdot \nabla_j W_{ji}\end{aligned} \quad (4.24)$$

### 4.4. Remarks of heat conduction and physical viscosity

Consider all the heat conduction happened in $\Omega(t)$,

$$\begin{aligned}&\sum_{\Omega(t)} \left(\sum_{j \in H_i} \frac{\kappa_i}{\rho_i^2} \frac{T_{ij}}{r_{ij}^2 + \eta^2} \mathbf{r}_{ij} \cdot \nabla_i W_{ij}\, m_j - \sum_{j \in H_i'} \frac{\kappa_j}{\rho_j^2} \frac{T_{ji}}{r_{ji}^2 + \eta^2} \mathbf{r}_{ji} \cdot \nabla_j W_{ji}\, m_j\right) m_i \\ =& \sum_{\Omega(t)} \sum_{j \in H_i} \frac{\kappa_i}{\rho_i^2} \frac{T_{ij}}{r_{ij}^2 + \eta^2} \mathbf{r}_{ij} \cdot \nabla_i W_{ij}\, m_j\, m_i - \sum_{\Omega(t)} \sum_{j \in H_i'} \frac{\kappa_j}{\rho_j^2} \frac{T_{ji}}{r_{ji}^2 + \eta^2} \mathbf{r}_{ji} \cdot \nabla_j W_{ji}\, m_j\, m_i \\ =& \sum_{\Omega(t)} \sum_{j \in H_i} \frac{\kappa_i}{\rho_i^2} \frac{T_{ij}}{r_{ij}^2 + \eta^2} \mathbf{r}_{ij} \cdot \nabla_i W_{ij}\, m_j\, m_i - \sum_{\Omega(t)} \sum_{j \in H_i} \frac{\kappa_i}{\rho_i^2} \frac{T_{ij}}{r_{ij}^2 + \eta^2} \mathbf{r}_{ij} \cdot \nabla_i W_{ij}\, m_j\, m_i \\ =& 0\end{aligned} \quad (4.25)$$

in the second step, the double summation in dual-support are converted to the double summation in support. The physical meaning of Eq.(4.25) is that the energy due to heat conduction in arbitrary domain $\mathbf{\Omega}(t)$ is conservative.

Similarly, consider the energy change in $\mathbf{\Omega}$ due to physical viscosity,

$$\sum_{\mathbf{\Omega}(t)} \left( -\frac{1}{2} \sum_{j \in H_i} \frac{\mu_i}{\rho_i^2} \frac{\mathbf{r}_{ij} \cdot \nabla_i W_{ij}}{r_{ij}^2 + \eta^2} v_{ij}^2 m_j - \frac{1}{2} \sum_{j \in H_i'} \frac{\mu_j}{\rho_j^2} \frac{\mathbf{r}_{ji} \cdot \nabla_j W_{ji}}{r_{ji}^2 + \eta^2} v_{ji}^2 m_j \right) m_i$$
$$= \sum_{\mathbf{\Omega}(t)} \left( -\sum_{j \in H_i} m_j \frac{\mu_i}{\rho_i^2} \frac{\mathbf{r}_{ij} \cdot \nabla_i W_{ij}}{r_{ij}^2 + \eta^2} v_{ij}^2 \right) m_i \qquad (4.26)$$

As $-\frac{\partial W_{ij}}{\partial r_{ij}} \geq 0$, Eq.(4.26) indicates that the physical viscosity is always transforming the kinetic energy into internal energy.

## 5. Conservation of basic laws

The conservation laws include the mass conservation, linear momentum conservation, angular momentum conservation and energy conservation. In this section, the conservation laws in DS-SPH are discussed. Since the set of equations is in Lagrangian representation, the total mass is conserved naturally.

### 5.1. Conservation of linear momentum

The momentum in domain $\mathbf{\Omega}(t)$ in the absence of external forces is given as

$$P(t) = \int_{\mathbf{\Omega}(t)} \rho_i \mathbf{v}_i \mathrm{d}V_i$$

The conservation of momentum requires

$$\frac{\mathrm{d}P}{\mathrm{d}t} = \int_{\mathbf{\Omega}(t)} \rho_i \frac{\mathrm{d}\mathbf{v}_i}{\mathrm{d}t} \mathrm{d}V_i = \mathbf{0} \qquad (5.1)$$

Or the discrete form

$$\frac{\mathrm{d}P}{\mathrm{d}t} = \sum_{\mathbf{\Omega}(t)} m_i \frac{\mathrm{d}\mathbf{v}_i}{\mathrm{d}t} = \mathbf{0} \qquad (5.2)$$

The variation of momentum is derived as

$$\begin{aligned}
\frac{\mathrm{d}P}{\mathrm{d}t} &= \sum_{\mathbf{\Omega}(t)} m_i \frac{\mathrm{d}\mathbf{v}_i}{\mathrm{d}t} \\
&= \sum_{\mathbf{\Omega}(t)} m_i \left( \sum_{j \in H_i} m_j \frac{\boldsymbol{\sigma}_i}{\rho_i^2} \cdot \nabla_i W_{ij} - \sum_{j \in H_i'} m_j \frac{\boldsymbol{\sigma}_j}{\rho_j^2} \cdot \nabla_j W_{ji} \right) \\
&= \sum_{\mathbf{\Omega}(t)} \sum_{j \in H_i} m_i m_j \frac{\boldsymbol{\sigma}_i}{\rho_i^2} \cdot \nabla_i W_{ij} - \sum_{\mathbf{\Omega}(t)} \sum_{j \in H_i'} m_i m_j \frac{\boldsymbol{\sigma}_j}{\rho_j^2} \cdot \nabla_j W_{ji} \\
&= \sum_{\mathbf{\Omega}(t)} \sum_{j \in H_i} m_i m_j \frac{\boldsymbol{\sigma}_i}{\rho_i^2} \cdot \nabla_i W_{ij} - \sum_{\mathbf{\Omega}(t)} \sum_{j \in H_i} m_i m_j \frac{\boldsymbol{\sigma}_i}{\rho_i^2} \cdot \nabla_i W_{ij} \\
&= \mathbf{0}
\end{aligned}$$

In the fourth step, the dual property of dual-support is used. Therefore, the conservation of momentum is satisfied.

*5.2. Conservation of angular momentum*

The angular momentum in domain $\mathbf{\Omega}(t)$ in the absence of external forces is given as

$$L(t) = \int_{\mathbf{\Omega}(t)} \mathbf{r}_i \times \rho_i \mathbf{v}_i \mathrm{d}V_i$$

The conservation of angular momentum requires

$$\begin{aligned}
\frac{\mathrm{d}L}{\mathrm{d}t} &= \int_{\mathbf{\Omega}(t)} \frac{\mathrm{d}}{\mathrm{d}t} \left( \mathbf{r}_i \times \rho_i \mathbf{v}_i \right) \mathrm{d}V_i \\
&= \int_{\mathbf{\Omega}(t)} \left( \rho_i \frac{\mathrm{d}\mathbf{r}_i}{\mathrm{d}t} \times \mathbf{v}_i + \rho_i \mathbf{r}_i \times \frac{\mathrm{d}\mathbf{v}_i}{\mathrm{d}t} \right) \mathrm{d}V_i \\
&= \int_{\mathbf{\Omega}(t)} \rho_i \mathbf{r}_i \times \frac{\mathrm{d}\mathbf{v}_i}{\mathrm{d}t} \mathrm{d}V_i \\
&= \mathbf{0}
\end{aligned} \quad (5.3)$$

Or the discrete form

$$\frac{\mathrm{d}L}{\mathrm{d}t} = \sum_{\mathbf{\Omega}(t)} m_i \mathbf{r}_i \times \frac{\mathrm{d}\mathbf{v}_i}{\mathrm{d}t} = \mathbf{0} \quad (5.4)$$

In the proof of angular momentum, we replace $\{\frac{1}{\rho_i^2}, \frac{1}{\rho_j^2}\}$ in momentum equation with $\frac{1}{\rho_i \rho_j}$. The variation of angular momentum is derived as

$$\frac{dL}{dt} = \sum_{\Omega(t)} m_i \mathbf{r}_i \times \frac{d\mathbf{v}_i}{dt}$$

$$= \sum_{\Omega(t)} m_i \mathbf{r}_i \times \left( \sum_{j \in H_i} m_j \frac{\boldsymbol{\sigma}_i}{\rho_i \rho_j} \cdot \nabla_i W_{ij} - \sum_{j \in H_i'} m_j \frac{\boldsymbol{\sigma}_j}{\rho_i \rho_j} \cdot \nabla_j W_{ji} \right)$$

$$= \sum_{\Omega(t)} \sum_{j \in H_i} m_i m_j \mathbf{r}_i \times \left( \frac{\boldsymbol{\sigma}_i}{\rho_i \rho_j} \cdot \nabla_i W_{ij} \right) - \sum_{\Omega(t)} \sum_{j \in H_i'} m_i m_j \mathbf{r}_i \times \left( \frac{\boldsymbol{\sigma}_j}{\rho_i \rho_j} \cdot \nabla_j W_{ji} \right)$$

$$= \sum_{\Omega(t)} \sum_{j \in H_i} m_i m_j \mathbf{r}_i \times \left( \frac{\boldsymbol{\sigma}_i}{\rho_i \rho_j} \cdot \nabla_i W_{ij} \right) - \sum_{\Omega(t)} \sum_{j \in H_i} m_i m_j \mathbf{r}_j \times \left( \frac{\boldsymbol{\sigma}_i}{\rho_i \rho_j} \cdot \nabla_i W_{ij} \right)$$

$$= \sum_{\Omega(t)} \sum_{j \in H_i} m_i m_j \mathbf{r}_{ij} \times \left( \frac{\boldsymbol{\sigma}_i}{\rho_i \rho_j} \cdot \nabla_i W_{ij} \right)$$

$$= \sum_{\Omega(t)} \sum_{j \in H_i} \mathbf{r}_{ij} \times (\boldsymbol{\sigma}_i \cdot \nabla_i W_{ij}) \Delta V_j \Delta V_i \tag{5.5}$$

In the fourth step, the dual property of dual-support is used. Temporally, for the sake of simplicity, we abbreviate $H_i \to H$, $\Delta V_j \to \Delta V$, $\mathbf{r}_{ij} \to \mathbf{r}$, $W_{ij} \to W$ and $\boldsymbol{\sigma}_i \to \boldsymbol{\sigma}$. Let capital letters $I, J, K, L$ be the dimensional index.

$$\sum_{j \in H_i} \mathbf{r}_{ij} \times (\boldsymbol{\sigma}_i \cdot \nabla_i W_{ij}) \Delta V_j$$

$$= \sum_H \mathbf{r} \times (\boldsymbol{\sigma} \cdot \nabla W) \Delta V$$

$$= \sum_H r_I \mathbf{e}_I \times (\sigma_{JK} \mathbf{e}_J \otimes \mathbf{e}_K \cdot W_{,L} \mathbf{e}_L) \Delta V$$

$$= \epsilon_{IJL} \sigma_{JK} \underbrace{\sum_H r_I W_{,K} \Delta V}_{=\delta_{IK}} \mathbf{e}_L$$

$$= \epsilon_{IJL} \sigma_{JI} \mathbf{e}_L = \mathbf{0}$$

where $\epsilon_{IJL}$ is the permutation symbol and $r_I$ refers to the I-th component of $\mathbf{r}$, $\nabla W$ is the function with gradient correction, and the symmetry of the Cauchy stress tensor $\boldsymbol{\sigma}$ and the completeness of $\nabla_i W_{ij}$ are used. Hence, Eq.(5.5) is zero, and the angular momentum is conserved when the linear completeness is satisfied. Obviously, the angular momentum is satisfied for continuous form since the kernel function without correction possesses the property of completeness.

Note that if $\boldsymbol{\sigma}$ is isotropic, e.g. inviscid fluid, then for kernel function $\nabla_i W_{ij}$ without gradient correction, $\boldsymbol{\sigma}_i \cdot \nabla_i W_{ij}$ in Eq.(5.5) is co-linear with the vector $\mathbf{r}_{ij}$; hence, $\mathbf{r}_{ij} \times (\boldsymbol{\sigma}_i \cdot \nabla_i W_{ij}) = 0$, in this case, the angular momentum is conserved even if the completeness of $\nabla_i W_{ij}$ is not satisfied.

*5.3. Conservation of energy*

The energy in domain $\mathbf{\Omega}(t)$ in the absence of external forces is given as

$$E(t) = \int_{\mathbf{\Omega}(t)} \rho_i e_i \mathrm{d}V_i$$

The conservation of energy requires

$$\frac{\mathrm{d}E}{\mathrm{d}t} = \int_{\mathbf{\Omega}(t)} \rho_i \frac{\mathrm{d}e_i}{\mathrm{d}t} \mathrm{d}V_i = 0$$

Or the discrete form

$$\frac{\mathrm{d}E}{\mathrm{d}t} = \sum_{\mathbf{\Omega}(t)} m_i \frac{\mathrm{d}e_i}{\mathrm{d}t} = 0 \tag{5.6}$$

The variation of specific energy is derived as

$$\begin{aligned}
\frac{\mathrm{d}E}{\mathrm{d}t} &= \sum_{\mathbf{\Omega}(t)} m_i \frac{\mathrm{d}e_i}{\mathrm{d}t} \\
&= \sum_{\mathbf{\Omega}(t)} \sum_{j \in H_i} m_i m_j \frac{\boldsymbol{\sigma}_i}{\rho_i^2} : (\mathbf{v}_j \otimes \nabla_i W_{ij}) - \sum_{\mathbf{\Omega}(t)} \sum_{j \in H_i'} m_i m_j \frac{\boldsymbol{\sigma}_j}{\rho_j^2} : (\mathbf{v}_i \otimes \nabla_j W_{ji}) \\
&= \sum_{\mathbf{\Omega}(t)} \sum_{j \in H_i} m_i m_j \frac{\boldsymbol{\sigma}_i}{\rho_i^2} : (\mathbf{v}_j \otimes \nabla_i W_{ij}) - \sum_{\mathbf{\Omega}(t)} \sum_{j \in H_i} m_i m_j \frac{\boldsymbol{\sigma}_i}{\rho_i^2} : (\mathbf{v}_j \otimes \nabla_i W_{ij}) \\
&= 0
\end{aligned} \tag{5.7}$$

In the third step, the dual property of dual-support is used. So the specific energy is conserved. It is worth mentioning that the proof of basic laws can be obtained as well with Eq.(4.19), Eq.(4.20), Eq.(4.22) and Eq.(4.23).

## 6. The implementation of DS-SPH

At each step, the values at step $t + \Delta t$ are calculated based on the known variables as $\mathbf{r}_i^t, \mathbf{v}_i^t, \mathbf{f}_i^t, \rho_i^t, \epsilon_i^t, \dot{\rho}_i^t, \dot{\epsilon}_i^t$. Many numerical schemes can be used to integrate the SPH formulation, e.g. the Leapfrog prediction-correction scheme and the Verlet-velocity scheme [38]. When the energy equation is considered, the Leapfrog prediction-correction scheme is preferred. The Leapfrog prediction-correction scheme comprises three steps as below

1. Prediction step

$$\mathbf{r}_i^{t+\Delta t} = \mathbf{r}_i^t + \Delta t \mathbf{v}_i^t + \frac{\Delta t^2}{2m_i} \mathbf{f}_i^t$$

$$\mathbf{v}_i^p = \mathbf{v}_i^t + \frac{\Delta t}{m_i} \mathbf{f}_i^t$$

$$\rho_i^p = \rho_i^t + \Delta t \dot{\rho}_i^t$$

$$\epsilon_i^p = \epsilon_i^t + \Delta t \dot{\epsilon}_i^t$$

2. Calculate all forces and all derivatives, $\mathbf{f}_i^{t+\Delta t}, \dot{\rho}_i^{t+\Delta t}, \dot{\epsilon}_i^{t+\Delta t}$ based on $\mathbf{r}_i^{t+\Delta t}, \mathbf{v}_i^p, \rho_i^p, \epsilon_i^p$.
3. Correction step

$$\mathbf{v}_i^{t+\Delta t} = \mathbf{v}_i^t + \frac{\Delta t}{2m_i}(\mathbf{f}_i^t + \mathbf{f}_i^{t+\Delta t})$$

$$\rho_i^{t+\Delta t} = \rho_i^t + \frac{\Delta t}{2}(\dot{\rho}_i^t + \dot{\rho}_i^{t+\Delta t})$$

$$\epsilon_i^{t+\Delta t} = \epsilon_i^t + \frac{\Delta t}{2}(\dot{\epsilon}_i^t + \dot{\epsilon}_i^{t+\Delta t})$$

During the second step of the scheme, the DS-SPH calculates of derivatives in a special way. Within the framework of DS-SPH, for any particle $j \in H_i$, we calculate the force $\mathbf{f}_{ij}$ for pair $ij$ and add it to particle $i$. Note that the vector $\mathbf{f}_{ij}$ considers two particles' mass, thus whose unit is force. Meanwhile, $j \in H_i \Leftrightarrow i \in H_j'$, we add $-\mathbf{f}_{ij}$ to particle $j$, by which one term in particle $j$'s dual-support is calculated. When summing for other particles, the forces from $H_i'$ are automatically done. In this sense, the force from one particle's support domain is reusable for the other particle's dual-support domain. We can see the dual-support for any particle is not stored but is inferred from the supports, and the forces from dual-support are automatically done.

To illustrate the summing process of forces, Eq.(4.19) is chosen as an example. When looping any particle $j \in H_i$, we add force $\mathbf{f}_{ij} = -m_i m_j (\frac{p_i}{\rho_i^2} + \frac{\Pi_{ij}}{2})\nabla_i W_{ij}$ to particle $i$; since $j \in H_i \Leftrightarrow i \in H_j'$, we add $-\mathbf{f}_{ij}$ to particle $j$. The case for the traditional SPH is different. When $j \in H_i$, we calculate $-m_i m_j(\frac{p_i}{\rho_i^2} + \frac{p_j}{\rho_j^2} + \Pi_{ij})\nabla_i W_{ij}$ for particle $i$; when $i \in H_j$, we also calculate $-m_i m_j(\frac{p_i}{\rho_i^2} + \frac{p_j}{\rho_j^2} + \Pi_{ji})\nabla_j W_{ji}$. Comparing the two force-summing methods, the DS-SPH not only reduces the computational cost of force for pair $ij$ but also conserves the basic laws exactly when variable supports are employed. Any particle in DS-SPH just concentrates on its own kernel function and support, while the kernel average method must consider two particles' kernel functions. When the radii of all supports are the same, the DS-SPH degenerates to the traditional constant support SPH.

When it is related to vary smoothing length, the concept of support and dual-support provides great flexibility. Eq.(2.20) indicates that the smoothing length is inversely proportional to density. Let $\Delta x_i$ denote particle $i$'s size. Without loss of generality, we assume that the shape of the particle is a cube in 3D or square in 2D. Then $\Delta x_i = \sqrt[d]{m_i/\rho_i}$, where $d$ is the number of dimensions. Since each particle's mass is fixed, $\Delta x_i$ decreases with density $\rho_i$ increased. At every $m(\approx 50)$ steps, we update the support radius based on the new estimated particle size by

$$h_i = n\Delta x_i = n\sqrt[d]{m_i/\rho_i} \tag{6.1}$$

where $n(\approx 1.5 \sim 3)$ is a global constant. It can be seen that the smoothing length decreases with the particle's density increasing; the low(high) density, which indicates the relatively sparse(dense) neighbors, causes the support domain to expand(shrink) adaptively.

# 7. Numerical examples

## 7.1. 1D heat conduction

Consider a 1d bar of with length $L$ =50 cm. The bar is discretized with particles with a particle spacing of $\Delta x$ =1cm or $\Delta x$ =0.5cm; the heat diffusion coefficient is $\alpha = 1.0 \times 10^{-4} m^2 s^{-1}$. The left half is assigned an internal energy of $e_l^0 = 1$ J/m, the right half $e_r^0 = 2$ J/m. The total energy in the 1d bar is $E_{total}$ =0.75 J. There is neither potential nor kinetic energy present in this simulation, so that energy conservation can be monitored by tracking the total internal energy. The energy profile is compared to an analytic solution after 4.0 s. This example was calculated with 2 SPH formulations, i.e. conventional SPH and dual-support SPH.

$$e(x,t) = \frac{e_r^0 + e_l^0}{2} + \frac{e_r^0 - e_l^0}{2}\mathrm{erf}\left(\frac{x-x_c}{\sqrt{4\alpha t}}\right) \tag{7.1}$$

Three cases as shown in Table 1 were considered. The first case is modeled by conventional SPH with constant smoothing length. The second case is simulated by conventional SPH with variable smoothing lengths but without additional treatment. The third case is implemented with our dual-support SPH. In order to test the influence of the transition of smoothing length, two particle spacings $\Delta x$ =1cm, $\Delta x$ =0.5cm in Case II and Case III were employed, more specific, the particle spacing in the interval of $x = 0.3L - 0.5L$ is $\Delta x$ =0.5cm.

The $L_2$ error in internal energy is given by

$$\|\mathrm{err}\|_{L_2} = \frac{\|\mathbf{e}^h - \mathbf{e}_{\mathrm{analytic}}\|}{\|\mathbf{e}_{\mathrm{analytic}}\|}, \tag{7.2}$$

with

$$\|\mathbf{e}\| = \left(\int_0^L \mathbf{e}\cdot\mathbf{e}\,dx\right)^{\frac{1}{2}}.$$

| Case | $\Delta x$ | $10^2 \cdot h$ | Particle numbers |
|------|-----------|----------------|------------------|
| I    | 0.01      | 3              | 100              |
| II   | 0.01,0.005 | 3/1.5         | 118              |
| III  | 0.01,0.005 | 3/1.5         | 118              |

Table 1: The parameters of three cases

The numerical results agreed well with theoretical solution, as shown in Table 2. The error of the conventional SPH formulation with varying smoothing length is roughly twice as high as the error with the dual-support SPH version.

| Case | $\Delta E/E_{total}$ | $\|\text{err}\|_{L_2}$ | $[e/e_{analytical}]_{max}$ |
|------|------|--------|---------|
| I    | -1.18E-15 | 0.0072 | -2.40% |
| II   | 3.38E-3   | 0.0115 | -5.15% |
| III  | -1.18E-15 | 0.0061 | -2.89% |

Table 2: The results of three cases for $t = 4$s

## 7.2. Sod shock wave tube

The sod shock tube is a good numerical benchmark problem and has been comprehensively studied by [39, 40, 41], to name a few. The shock-tube is a long straight tube filled with gas, which is separated by a membrane into two parts of different pressures and densities. The initial conditions are given by

$$x \leq 0 \quad \rho = 1 \quad v = 0 \quad e = 1.5 \quad p = 1 \quad \Delta x = 0.00125$$
$$x > 0 \quad \rho = 0.125 \quad v = 0 \quad e = 1.2 \quad p = 0.1 \quad \Delta x = 0.01$$

where $\rho$, $p$, $e$, and $v$ are the density, pressure, internal energy, and velocity of the gas, respectively. $\Delta x$ is the initial particle spacing. Two cases were tested to compare the conventional SPH with variable smoothing lengths [22] and the dual-support SPH. In case I, the smoothing length is determined by solving Eq.(2.20) with an iteration method. In Case II, the initial smooth size for each particle is set as 2 times of the particle size. The artificial viscosity is calculated by Eq.(4.17), where $\alpha = 1, \beta = 2$. The density is calculated by Eq.(4.8). The relation between the internal energy and pressure is given by

$$p = (\gamma - 1)\rho e \tag{7.3}$$

where $\gamma = 5/3$. The leapfrog prediction-correction method is used to integrate the process.

The velocity profile, density profile, pressure profile and energy profile at $t = 0.2$ s are shown in Fig.3 and Fig.4. It can be seen that the solution obtained by DS-SPH is accurate compared with that by the smoothing length variable SPH [22].

## 7.3. Water bubble

In this section the flow simulation of an elliptical water bubble is presented. The example is to show that the total linear, angular momentum, and energy for dual-support SPH are well preserved in the absence of external forces. This example is tested by two formulations, one is the Eq.(4.10) with only one support for each particle (namely, single support SPH), the other is Eq.(4.13) (dual-support SPH). The gradient correction of kernel function (cubic-spline kernel) is employed in two formulations. Although the energy equation is not directly involved in this problem, in order to track the total energy, Eq.(4.12) and Eq.(4.15) are solved. The total energy comprises three two parts, the gravitational potential energy, the internal energy and the kinetic energy. The continuity equation Eq.(4.1) is calculated directly by the divergence of velocity. The velocity gradient is calculated by

$$\nabla \otimes \mathbf{v}_i = -\sum_{j \in H_i} \frac{m_j}{\rho_j} \mathbf{v}_{ij} \otimes \tilde{\nabla}_i W_{ij}. \tag{7.4}$$

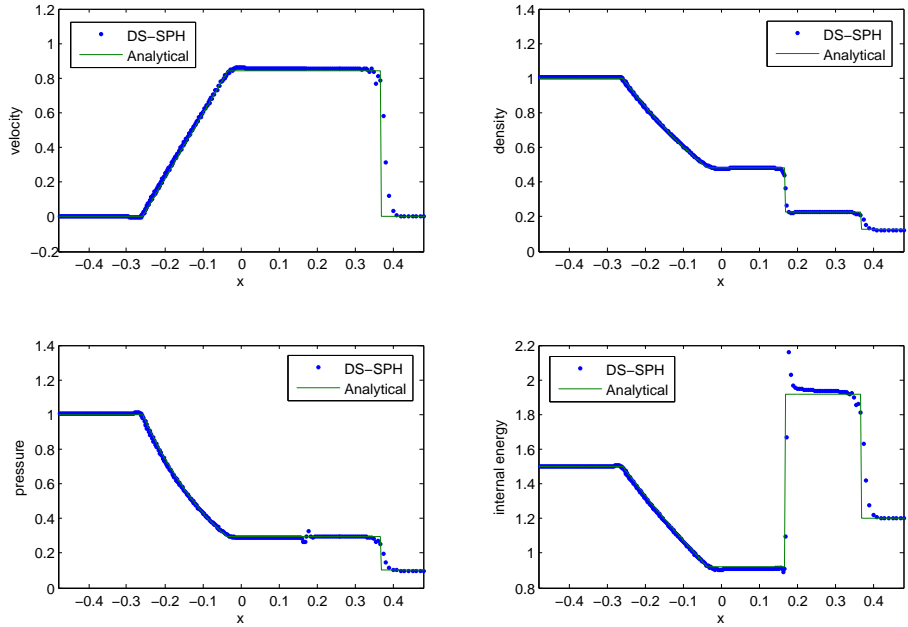

Figure 3: The results of 1D sod shock tube by DS-SPH at time $t = 0.2$s

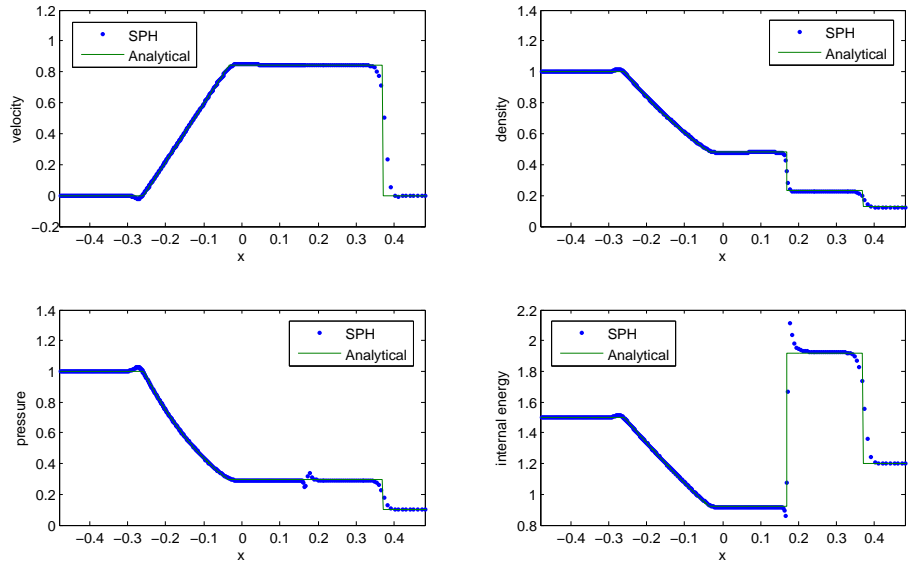

Figure 4: The results of 1D sod shock tube with correction term $\nabla h$ [22] at time $t = 0.2$s

The geometry of the bubble is a circle of 1 m radius without external forces but with initial velocity field of $(-100x, 100y)$ m/s. The bubble is discretized with 1979 particles, whose distribution is shown in Fig.5. The cell's volume represents the particle's volume. In order to test the influence of variable support sizes on the conservation laws, the upright part of the bubble is discretized with small particles. The support size is selected as two times of the particle size, i.e. $h_i = 2\Delta x_i$. The particle size is estimated by assuming the shape of square.

The total stress is calculate by Eq.(4.6). The pressure is calculated by the following equation of state for water [10],

$$p = p_0((\frac{\rho}{\rho_0})^\gamma - 1), \ c^2(\rho) = \frac{\partial p}{\partial \rho} = c_0^2 (\rho/\rho_0)^{\gamma-1}, \qquad (7.5)$$

where $\gamma$ is a constant, and $\gamma = 7$ is used in most circumstances. $\rho_0$ is the reference density, $c_0$ is the sound speed at the reference density. $p_0 = c_0^2 \rho_0 / \gamma$ is the artificial bulk modulus [33]. In this example, the artificial bulk modulus is $p_0 = 285.714$ MPa, dynamic viscosity $\mu = 0.5$ kg m$^{-1}$s$^{-1}$, initial density $\rho_0 = 10^3$ kg/m$^3$. Within the simulations, the shape of the bubble should remain elliptical, the value of $ab$ (semi-major axis × semi-minor axis) should remain constant. The analytical solution of $b$ varying with time can be obtained as

$$\frac{\mathrm{d}b}{\mathrm{d}t} = -bB, \ \text{where} \ \frac{\mathrm{d}B}{\mathrm{d}t} = \frac{B^2(b^4 - w^4)}{b^4 + w^4} \qquad (7.6)$$

and $w$ is the initial value of $ab$.

Fig.(6) shows the geometry of the water bubble simulated by single support SPH and the dual-support SPH. Both results agreed well with the theoretical shape denoted by solid lines, between which there is no visible difference. In this case, the influence of variable support size is not so significant due to the gradual transition of support size. However, when tracking the total linear momentum (Fig.7), total angular momentum (Fig.8) and total energy (Fig.9), the results by single support SPH changed significantly with time, whereas that by dual-support SPH were well preserved.

*7.4. 2D Dam break over dry bed*

The Dam break experiment, which was described in reference [42], is a benchmark problem to test the accuracy of SPH code by Violeau and Issa [43] and Crespo ta al[44]. The tank is 4 m long, the initial volume of water is 1 m long and its height 2m, as shown in Fig. 10. The system is solved with a leapfrog prediction-correction scheme, using a cubic-spline kernel without kernel gradient correction, specular reflection boundary condition by Eq.(7.7), artificial viscosity, $\alpha = 1$, $\beta = 1$. The density is calculated by Eq.(4.9). Fluid particles were initially placed on a staggered grid with zero initial velocity. In order to employ a large time increment, the sound speed is set as 100 m/s, which is 10 times larger than the maximum flowing speed. The specular reflection boundary is given by

$$\mathbf{v}' = \mathbf{v} - 2(\mathbf{v} \cdot \mathbf{n})\mathbf{n} \ \text{if} \ \mathbf{v} \cdot \mathbf{n} < 0 \ , \qquad (7.7)$$

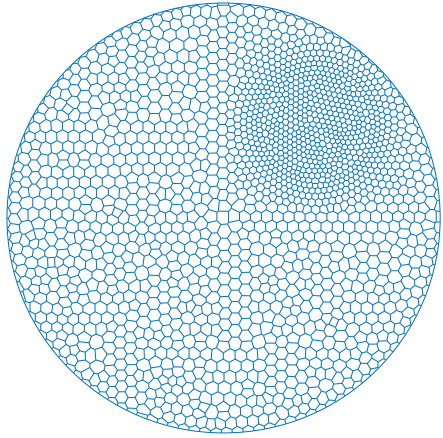

Figure 5: The particle discretization of water bubble. Each cell represents one particle.

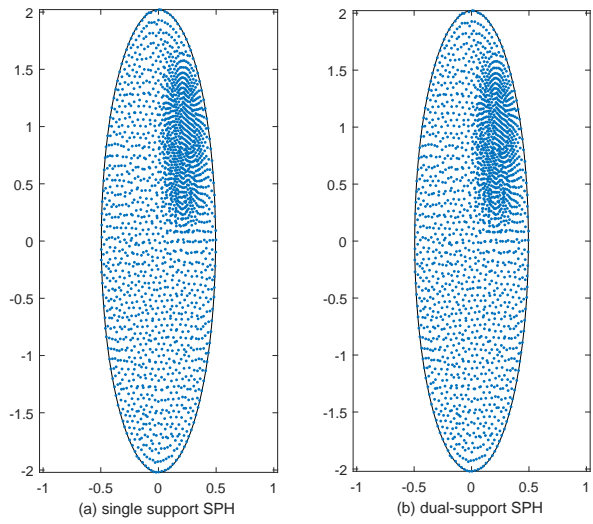

Figure 6: The geometry of water bubble at $t = 8 \times 10^{-3}$ s. The solid line represents the theoretical shape.

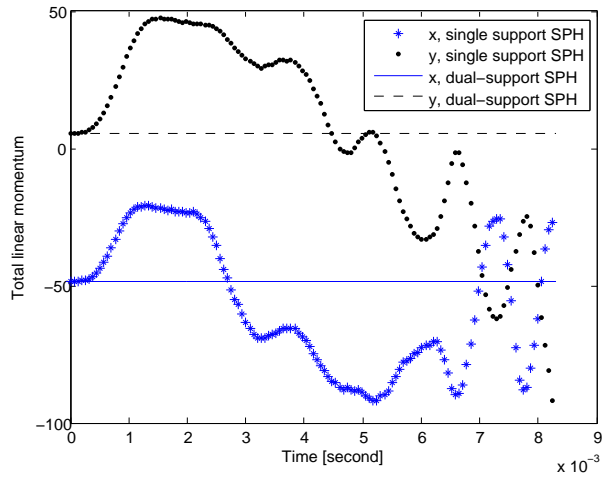

Figure 7: The total linear momentum against time.

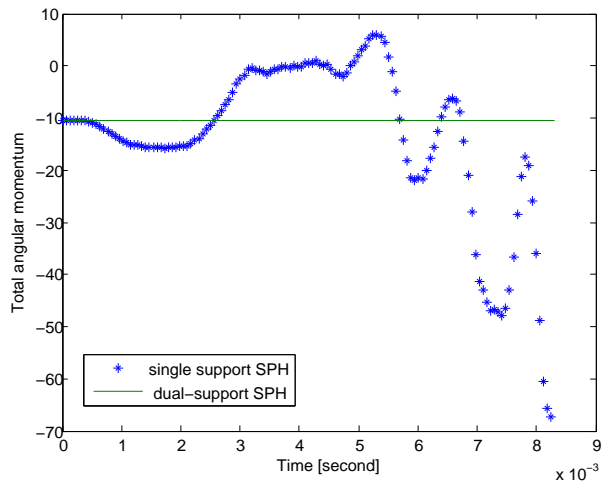

Figure 8: The total angular momentum against time.

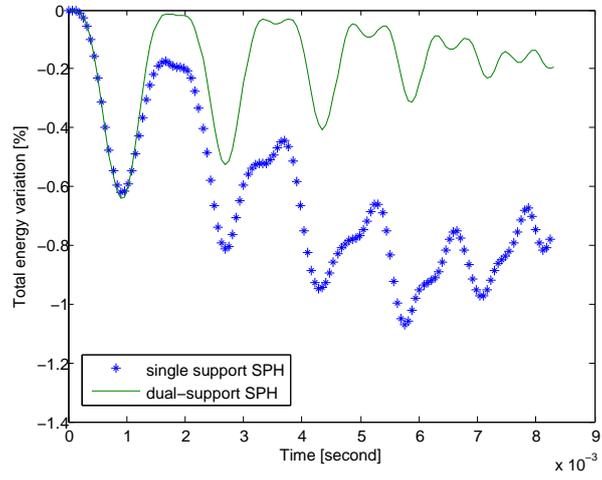

Figure 9: The total energy variation against time.

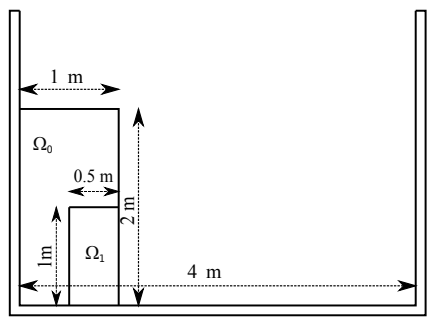

Figure 10: Initial configuration of the water column and the tank

where **v** denotes the velocity, **n** is the inward normal direction of the wall at that point. Note that the specular reflection boundary only changes the direction of the particle's velocity when the particle is approaching the boundary, thus the kinetic energy is not effected. Such good property enables us to track the global energy during the simulation. Although the equation of state given by Eq.(7.5) is irrelevant to the energy state, the energy equation Eq.(4.18) or Eq.(4.20) are considered in order to track the internal energy since the artificial viscosity converts the kinetic energy into the internal energy.

Three cases are run to shown the capabilities of the dual-support SPH. Case I is based on traditional single support SPH but with variable support size; the support size is proportional to the particle size; Case III uses constant support size during the simulation. The particle spacing of Case I (5,600 particles) and Case II (5,600 particles) is $\Delta x = 2.5 \times 10^{-2}$ m in $\Omega_0$, $\Delta x = 1.25 \times 10^{-2}$ m in $\Omega_1$, whereas only one particle spacing $\Delta x = 1.25 \times 10^{-2}$ in Case III (12,800 particles). The smoothing length is set as $h_i = 2\Delta x_i$. The support size is updated with Eq.(6.1) at every 200 steps. The only difference between Case I and Case II is that Case II employed the dual-support SPH formulation. The Case III based on traditional constant support SPH is served as the reference. There is a sharp change of support size in the interface of $\Omega_1$ and $\Omega_2$ for Case I and Case II. Such artificial transition deteriorates the traditional single support SPH, while dual-support SPH can reduce the bad effect.

As shown in Fig.11, the toe velocity of Case II and Case III agreed well with experimental data, whereas that of Case I was affected by sharp transition of support sizes.

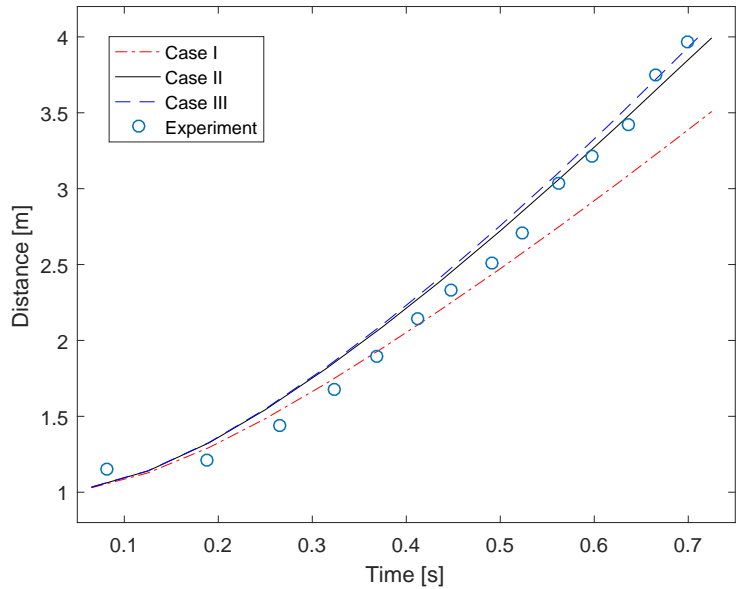

Figure 11: The evolution of dam front toe in experiment [42] and three cases.

The fluid domain is marked with four colors so that the deformation of the interfaces can be tracked. The deformation of water column for three cases at different time is shown in Fig.12 and Fig.13. For case I, the blue zone pushed the red zone and caused the simulation to deviate far away from the reference results given by Case III; the interfaces became irregular and the blue zone spanned over the bottom. The results of Case II agreed well the that by Case III; the interfaces were continuous and smooth in all steps. Hence, the dual-support SPH can reduce the adverse effect of variable support sizes to a minimum. The main difference between Case II and Case III is due to the degrees of freedom.

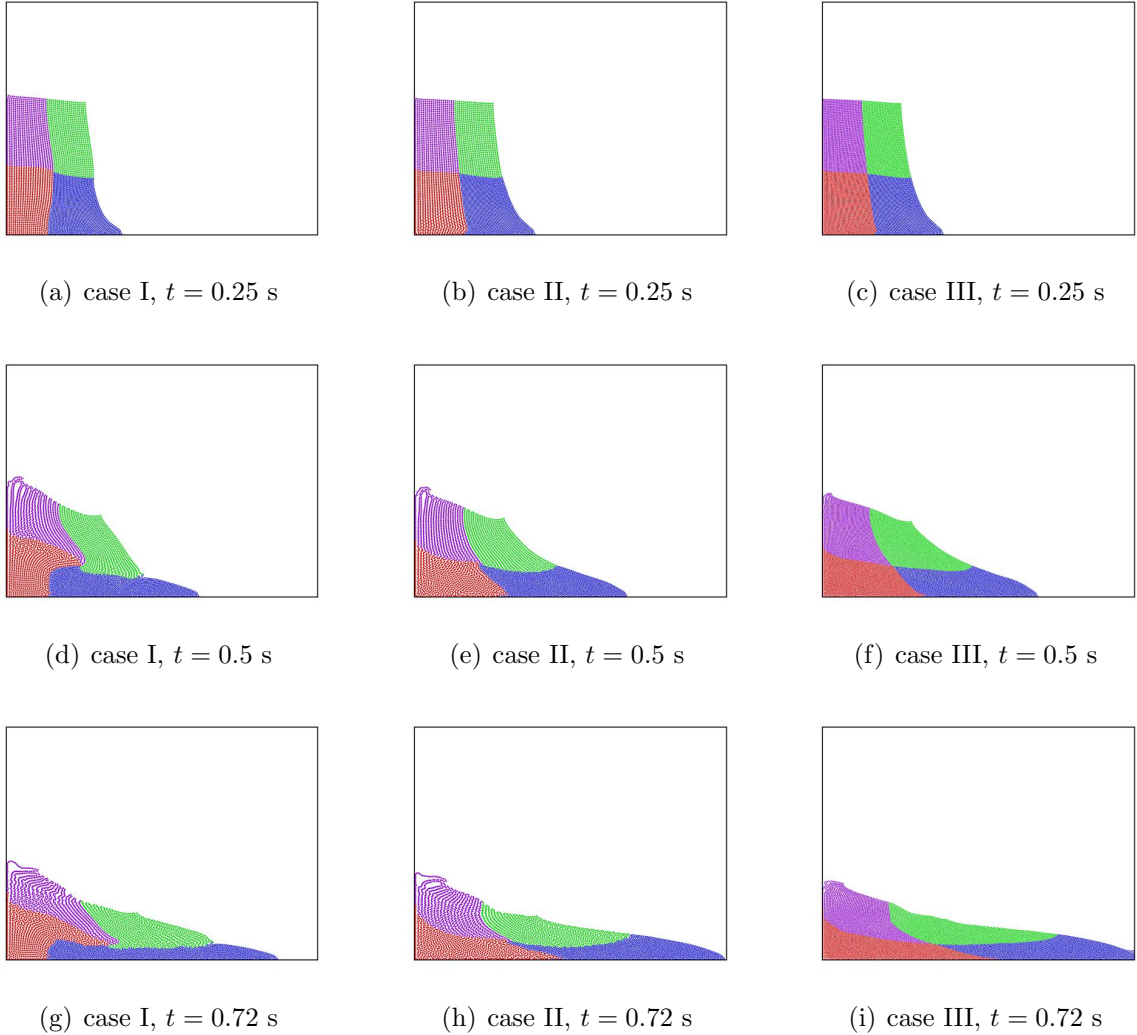

Figure 12: The profile of particles at different time

When the front toe hit the right wall, for Case II and Case III, the maximal density variation for front toe is smaller than 2%, as shown in Fig. 14, while the density for other parts of the fluid is very close to the initial density.

The total energy for three cases were compared in Fig.15. Fig.15(a) shows that the total energy of Case I increased over time, which indicates that there is spurious force doing work. In fact, the spurious force is the unbalance force interaction between two particles with different support sizes. For Case II and Case III, the total energy variation is less than 4 %, which is due to the numerical integration method. Therefore, the dual-support SPH

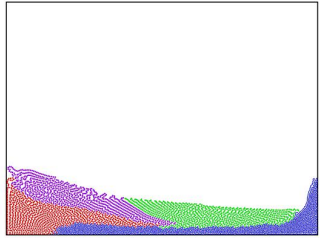
(a) case I, $t = 1.0$ s

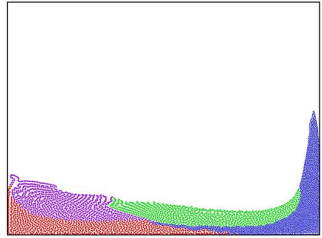
(b) case II, $t = 1.0$ s

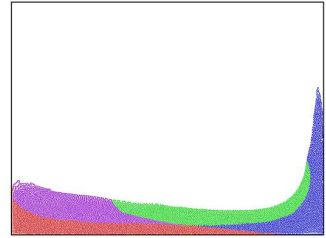
(c) case III, $t = 1.0$ s

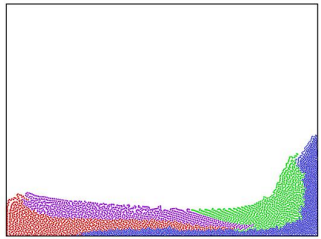
(d) case I, $t = 1.38$ s

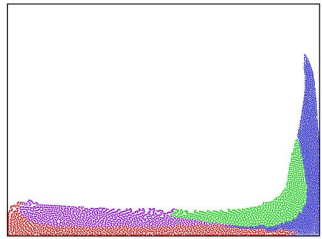
(e) case II, $t = 1.38$ s

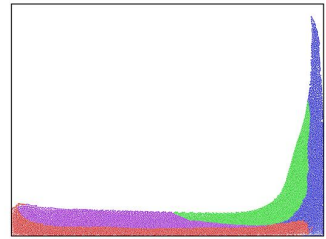
(f) case III, $t = 1.38$ s

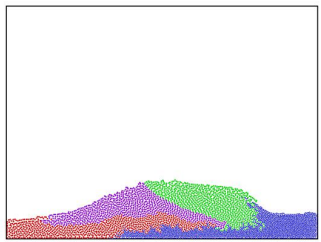
(g) case I, $t = 2.5$ s

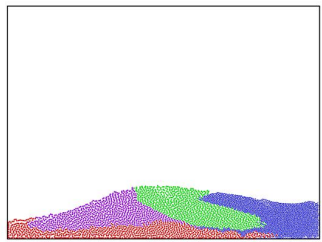
(h) case II, $t = 2.5$ s

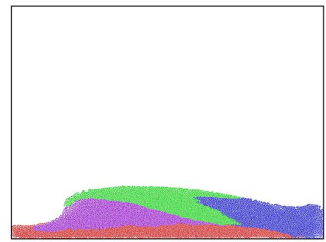
(i) case III, $t = 2.5$ s

Figure 13: The profile of particles at different time

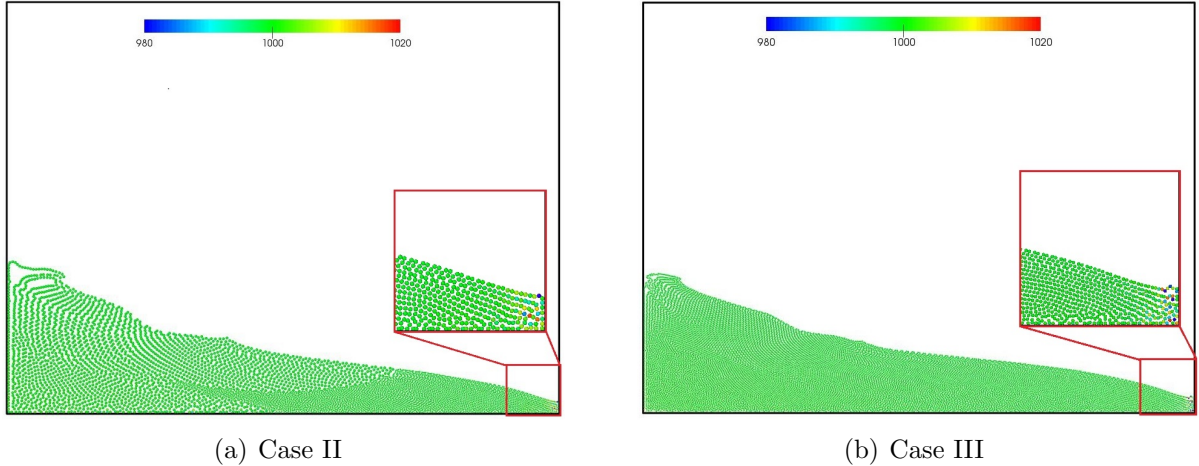

(a) Case II　　　　　　　　　　　　　　　(b) Case III

Figure 14: The density contour at different time

formulation conserves the global energy in the absence of external force doing work.

## 8. Conclusions

In this paper, we have introduced the dual-support domain, the dual item of support domain, based on which the conventional SPH was reformulated into dual-support SPH. This formulation enables the conservations of momentum, angular momentum and energy at the same time when variable supports are employed. In DS-SPH, the change of smoothing length is achieved with ease and the modified coefficients in Eq.(2.15) and Eq.(2.17) are eliminated. The implementation of DS-SPH shows that the force is only calculated once, the reaction force is calculated by adding negative symbol, thus reducing the computational cost compared with traditional SPH.

Four numerical examples were presented to validate the dual-support SPH. The first numerical example shows for heat diffusion problem the energy is conserved when variable supports are utilized. The second numerical example shows that the dual-support SPH can achieve good accuracy compared with other SPH with variable smoothing lengths. The third numerical example verifies that the conservations of basic laws for DS-SPH are well preserved while the traditional SPH not when variable smoothing lengths used. The last example shows traditional SPH is affected by the variable smoothing lengths and DS-SPH can eliminate the adverse effect.

The concept of dual-support facilitates the support-variable SPH formulation. To some extent, the concept of support and dual-support is similar to Newton's third law, considering the direct force and reaction force for paired particles. The proof of conservation is obtained easily with the aid of dual property of dual-support. It can be seen throughout the paper that there is few restrictions on the dual-support, therefore, the dual-support can be also applied in the SPH on solid or SPH on magnetohydrodynamics. The present method is also promising for multiscale analysis where the models with different length scales can be bridged by using different smooth length settings.

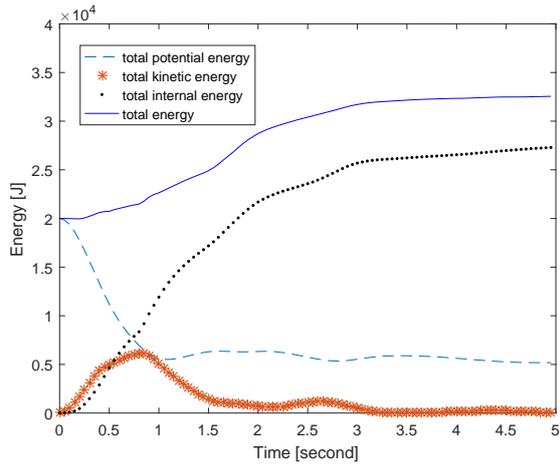
(a) Case I

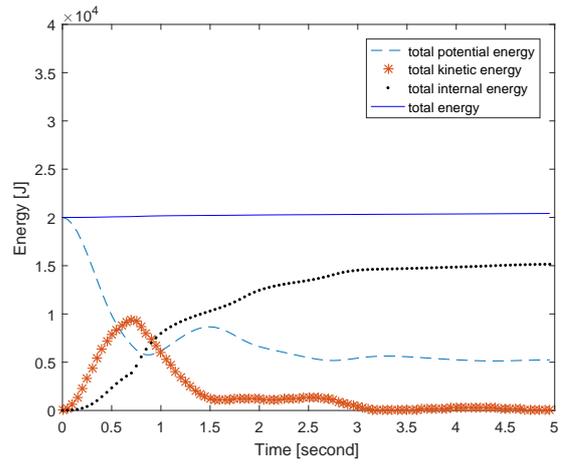
(b) Case II

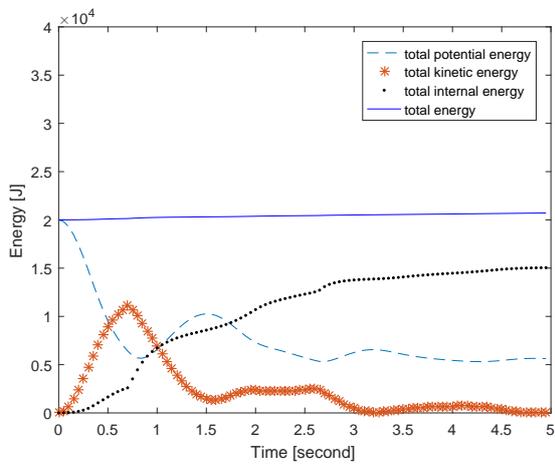
(c) Case III

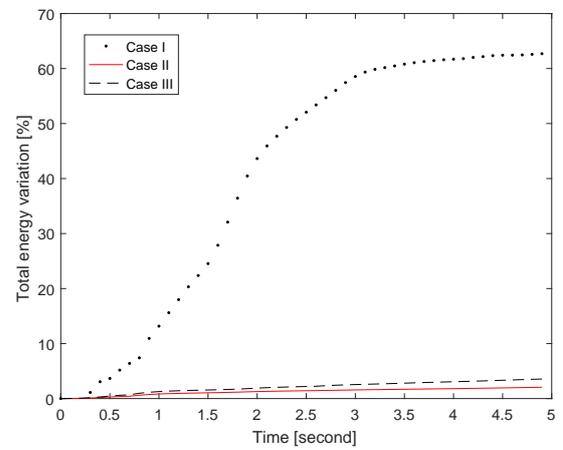
(d) Energy variation for 3 cases

Figure 15: The total energy at different time

**Acknowledgements**

The authors acknowledge the supports from FP7 Marie Curie Actions ITN-INSIST, and ERC-CoG( Computational Modeling and Design of Lithium-ion Batteries (COMBAT)), the National Basic Research Program of China (973 Program: 2011CB013800) and NSFC (51474157), the Ministry of Science and Technology of China (Grant No.SLDRCE14-B-28, SLDRCE14-B-31).

**Appendix A. The dual property of dual-support**

Let $\mathcal{F}(i,j)$ be any expression depend on two points $i,j$. The dual property of dual-support is that the double integral of the term $\mathcal{F}(i,j)$ in dual-support can be converted to the double integral of the term $\mathcal{F}(j,i)$ in support, as shown in Eq. (3.4) and Eq. (3.5). The key idea lies in that the term $\mathcal{F}(i,j)$ can be both interpreted in $H_i$ and $H_j'$.

$$\sum_{i \in \mathbf{\Omega}} \left( \sum_{j \in H_i'} \mathcal{F}(i,j) \, \Delta V_j \right) \Delta V_i = \sum_{i \in \mathbf{\Omega}} \left( \sum_{j \in H_i} \mathcal{F}(j,i) \, \Delta V_j \right) \Delta V_i \qquad \text{discrete form} \qquad (A.1)$$

$$\int_{i \in \mathbf{\Omega}} \int_{j \in H_i'} \mathcal{F}(i,j) \, \mathrm{d}V_j \mathrm{d}V_i = \int_{i \in \mathbf{\Omega}} \int_{j \in H_i} \mathcal{F}(j,i) \, \mathrm{d}V_j \mathrm{d}V_i \qquad \text{continuous form} \qquad (A.2)$$

**Proof:**
Let $\mathbf{\Omega}$ be discretized with $N$ voronoi tessellations (or other shape), as shown in Fig. A.16.

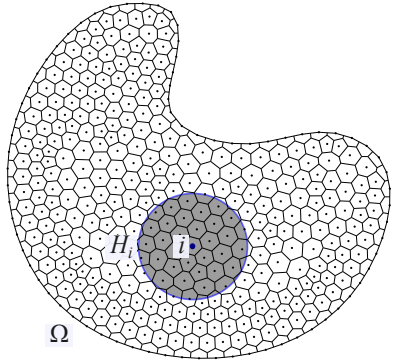

Figure A.16: The discretization of domain $\mathbf{\Omega}$

Each polygon is denoted with an index $i \in \{1, \cdots, N\}$, $\mathbf{r}_i$ is the coordinate for $i$'s center of gravity, $\Delta V_i$ is the volume associated to $i$, $H_i$ and $H_i'$ are $i$'s support and dual-support, respectively. So

$$\mathbf{\Omega} = \sum_{i=1}^{N} \Delta V_i$$

Consider the double summation of $\mathcal{F}(i,j)$ on $\boldsymbol{\Omega}$.

$$\sum_{\Delta V_i \in \Omega} \left( \sum_{\Delta V_j \in H'_i} \mathcal{F}(i,j) \Delta V_j \right) \Delta V_i$$

$$= \sum_{1 \leq i \leq N} \left( \sum_{j \in H'_i} \mathcal{F}(i,j) \Delta V_j \right) \Delta V_i$$

$$= \sum_{j \in H'_1} \mathcal{F}(1,j) \Delta V_j \Delta V_1 + \sum_{j \in H'_2} \mathcal{F}(2,j) \Delta V_j \Delta V_2 + \cdots + \sum_{j \in H'_N} \mathcal{F}(N,j) \Delta V_j \Delta V_N \quad \text{(A.3)}$$

In the third step, $j \in \{1, \cdots, N\}$ means $j$ belongs to that point's dual-support. Each term $\mathcal{F}(i,j) \Delta V_j \Delta V_i$ in $H'_i$ can be interpreted as term $\mathcal{F}(i,j) \Delta V_i \Delta V_j$ in $H_j$. Let us sum all terms in a way based on point $j$'s support $H_j$, where $j \in \{1, \cdots, N\}$

$$\sum_{j \in H'_1} \mathcal{F}(1,j) \Delta V_j \Delta V_1 + \sum_{j \in H'_2} \mathcal{F}(2,j) \Delta V_j \Delta V_2 + \cdots + \sum_{j \in H'_N} \mathcal{F}(N,j) \Delta V_j \Delta V_N$$

$$= \sum_{i \in H_1} \mathcal{F}(i,1) \Delta V_1 \Delta V_i + \sum_{i \in H_2} \mathcal{F}(i,2) \Delta V_2 \Delta V_i + \cdots + \sum_{i \in H_N} \mathcal{F}(i,N) \Delta V_N \Delta V_i \quad \text{(A.4)}$$

In the second step of Eq. (A.4), for example, $\sum_{i \in H_1} \mathcal{F}(i,1) \Delta V_1 \Delta V_i$ means gathering all terms $j = 1$ in

$$\sum_{j \in H'_2} \mathcal{F}(2,j) \Delta V_j \Delta V_2 + \sum_{j \in H'_3} \mathcal{F}(3,j) \Delta V_j \Delta V_3 + \cdots + \sum_{j \in H'_N} \mathcal{F}(N,j) \Delta V_j \Delta V_N$$

$$\sum_{i \in H_1} \mathcal{F}(i,1) \Delta V_1 \Delta V_i + \sum_{i \in H_2} \mathcal{F}(i,2) \Delta V_2 \Delta V_i + \cdots + \sum_{i \in H_N} \mathcal{F}(i,N) \Delta V_N \Delta V_i$$

$$= \sum_{j \in H_1} \mathcal{F}(j,1) \Delta V_j \Delta V_1 + \sum_{j \in H_2} \mathcal{F}(j,2) \Delta V_j \Delta V_2 + \cdots + \sum_{j \in H_N} \mathcal{F}(j,N) \Delta V_j \Delta V_N$$

$$= \sum_{1 \leq i \leq N} \left( \sum_{j \in H_i} \mathcal{F}(j,i) \Delta V_j \right) \Delta V_i \quad \text{(A.5)}$$

In the second step of Eq. (A.5), $i$ and $j$ is swapped. Eqs. (A.3-A.5) lead to

$$\sum_{1 \leq i \leq N} \left( \sum_{j \in H'_i} \mathcal{F}(i,j) \Delta V_j \right) \Delta V_i = \sum_{1 \leq i \leq N} \left( \sum_{j \in H_i} \mathcal{F}(j,i) \Delta V_j \right) \Delta V_i$$

When $N \to \infty$ so that $\Delta V_i \to 0$, we have

$$\lim_{N \to \infty} \sum_{1 \leq i \leq N} \left( \sum_{j \in H'_i} \mathcal{F}(i,j) \Delta V_j \right) \Delta V_i = \int_{i \in \Omega} \int_{j \in H'_i} \mathcal{F}(i,j) \mathrm{d} V_j \mathrm{d} V_i \quad \text{(A.6)}$$

Hence, the dual property of dual-support in the integral form is

$$\int_{i\in\Omega}\int_{j\in H'_i} \mathcal{F}(i,j)\, \mathrm{d}V_j \mathrm{d}V_i = \int_{i\in\Omega}\int_{j\in H_i} \mathcal{F}(j,i)\, \mathrm{d}V_j \mathrm{d}V_i \tag{A.7}$$

Eq. (A.7) means the double integral of the term in dual-support can be converted to the double integral of the term with $i$ and $j$ swapped in support.

Note that the domain is not necessary to be continuous, $\Delta V_i, \Delta V_j$ can be replaced with sparse particles with mass $m_i, m_j$, respectively. So, we get

$$\sum_{i\in\Omega}\sum_{j\in H'_i} m_i\, m_j\, \mathcal{F}(i,j) = \sum_{i\in\Omega}\sum_{j\in H_i} m_i\, m_j \mathcal{F}(j,i) \tag{A.8}$$

particles in SPH methods. *CMC-TECH SCIENCE PRESS-*, 5(3):173, 2007.